\documentclass[a4paper,11pt]{article}
\pdfoutput=1 

\usepackage{jheppub} 
\usepackage[utf8]{inputenc}
\usepackage{graphicx}
\usepackage{appendix}

\usepackage{natbib}
\usepackage{slashed}
\usepackage{makecell}
\title{ \LARGE Superradiant dark matter production from primordial black holes: Impact of multiple modes and gravitational wave emission}

\author[a]{Nayun Jia,}
\author[b]{Shou-Shan Bao,}
\author[a,\ast]{Chen Zhang,}\note[$\ast$]{Corresponding author.}
\author[b]{Hong Zhang}
\author[a,c,d,\ast]{and Xin Zhang}

\affiliation[a]{Liaoning Key Laboratory of Cosmology and Astrophysics, College of Sciences, Northeastern University, NO. 3-11, Wenhua Road, Heping District, Shenyang 110819, China}
\affiliation[b]{Institute of Frontier and Interdisciplinary Science, Key Laboratory of Particle Physics and Particle Irradiation (MOE), Shandong University, 72 Jimobinhai Road, Qingdao 266237, China}
\affiliation[c]{MOE Key Laboratory of Data Analytics and Optimization for Smart Industry, Northeastern University, NO. 3-11, Wenhua Road, Heping District, Shenyang 110819, China}
\affiliation[d]{National Frontiers Science Center for Industrial Intelligence and Systems Optimization, Northeastern University, NO. 3-11, Wenhua Road, Heping District, Shenyang 110819, China}

\emailAdd{nayun.jia@foxmail.com}
\emailAdd{ssbao@sdu.edu.cn}
\emailAdd{zhangchen2@mail.neu.edu.cn}
\emailAdd{hong.zhang@sdu.edu.cn}
\emailAdd{zhangxin@mail.neu.edu.cn}

\abstract{Rotating primordial black holes (PBHs) in the early universe can emit particles through superradiance, a process particularly efficient when the particle's Compton wavelength is comparable to the PBH's gravitational radius. Superradiance leads to an exponential growth of particle occupation numbers in gravitationally bound states. We present an analysis of heavy bosonic dark matter (DM) production through three gravitational mechanisms: Hawking radiation, superradiant instabilities, and ultraviolet (UV) freeze-in. We consider PBHs that evaporate before Big Bang Nucleosynthesis (BBN). For both scalar and vector DM, our analysis incorporates the evolution of a second superradiant mode. We demonstrate that the growth of a second superradiant mode causes the decay of the first mode, and thus the second mode cannot further enhance the DM abundance beyond that already achieved by the first mode. Our study also reveals that while superradiance generally enhances DM production, gravitational wave (GW) emission from the superradiant cloud may significantly modify this picture. For scalar DM, GW emission reduces the parameter space where superradiance effectively augments relic abundance. For vector DM, rapid GW emission from the superradiant cloud may yield relic abundances below those achieved through Hawking radiation alone.  These findings demonstrate that multiple-mode effect and GW emission play critical roles in modeling DM production from PBHs in the early universe.}

\begin{document}

\maketitle

\section{Introduction}\label{sec:intro}

The existence of dark matter (DM) is strongly supported by extensive gravitational evidence, including galactic rotation curves, gravitational lensing, and cosmic microwave background (CMB) anisotropies (e.g., refs.~\cite{vanAlbada:1984js,Treu:2004wt,DES:2021wwk,Planck:2018vyg,Cirelli:2024ssz,Profumo:2017hqp,Mambrini:2021cwd}). Yet, the nature of DM remains elusive. While most studies focus on DM candidates that interact with the Standard Model (SM) particles beyond gravity, such as weakly interacting massive particles (WIMPs) (e.g., refs.~\cite{Arcadi:2017kky,Roszkowski:2017nbc}), axions and axion-like particles (e.g., refs.~\cite{Marsh:2015xka,Irastorza:2018dyq}), or dark sectors coupled through portals (e.g., refs.~\cite{Elahi:2014fsa,Albouy:2022cin}), the hypothesis that DM interacts solely via gravity with the SM particles remains compelling. In this gravitationally minimal scenario, the key challenge is to account for the observed DM relic abundance using solely gravitational interactions in the early universe.

Primordial black holes (PBHs), hypothesized to arise from various origins, have been extensively investigated for their cosmological implications, potential to shed light on physics beyond the SM, and distinctive observational signatures~\cite{Ali-Haimoud:2017rtz,Bird:2022wvk,Zhang:2023tfv,Zhu:2023gmx,Carr:2021bzv,Zhang:2023zmb,Yang:2024vij,Yang:2025uvf,Yang:2024pfb,Zhao:2024yus,Escriva:2022duf,Carr:2020xqk,Liu:2021svg,He:2022amv,Pi:2021dft,Fu:2019ttf,Di:2017ndc,Yi:2020kmq,Gu:2022pbo,Cai:2018tuh,Wang:2016ana,Wang:2022nml,Huang:2023mwy,Tan:2022lbm,Friedlander:2022ttk}. In this work, instead of considering PBHs themselves as DM candidates, we focus on DM production from them. They offer an intriguing solution to the challenge of accounting for the observed DM relic abundance using solely gravitational interactions. Among the three gravitational mechanisms contributing to DM genesis---Hawking radiation, superradiant instabilities, and gravitational ultraviolet (UV) freeze-in---PBH is the essential ingredient of the first two processes.

Hawking radiation~\cite{Hawking:1975vcx},\footnote{In this work, Hawking radiation is treated based on the usual semiclassical results. Recent studies suggest the memory burden effect may
modify the conventional picture of Hawking radiation~\cite{Dvali:2018xpy,Dvali:2020wft,Dvali:2024hsb}, which we do not consider in this work
but leave it to future studies.} arising from quantum effects in curved spacetime, provides a universal mechanism for particle production that operates independently of any additional interactions beyond gravity. A comprehensive analysis in ref.~\cite{Cheek:2021odj} demonstrates that PBH evaporation can contribute significantly to DM abundance through two channels: direct DM production from Hawking radiation, and indirect production through the decay of additional unstable particle species initially emitted by evaporating PBHs.

PBHs can acquire angular momenta through various mechanisms, such as domain wall collapse~\cite{Eroshenko:2021sez}, gravitational collapse during a matter-dominated era~\cite{Harada:2017fjm,Kuhnel:2019zbc}, assembly of matter-like objects such as Q-balls or oscillons~\cite{Flores:2021tmc}, and emission of light scalar particles~\cite{Chambers:1997ai,Calza:2021czr,Calza:2023rjt}. The acquisition of angular momentum enables superradiant instabilities. For rotating (Kerr) PBHs, superradiance can extract rotational energy to form gravitationally bound particle clouds, providing an additional channel for DM production beyond Hawking radiation. The phenomenon of black hole (BH) superradiance has been extensively investigated in the literature (e.g., refs.~\cite{Arvanitaki:2009fg,Berti:2009kk,Arvanitaki:2010sy,Brito:2013wya,Brito:2014wla,
Arvanitaki:2014wva,Li:2014fna,Marsh:2015xka,Wang:2015fgp,Huang:2016qnk,Endlich:2016jgc,
Rosa:2017ury,Baryakhtar:2017ngi,Huang:2018qdl,Li:2019tns,Chen:2019fsq,Zhang:2020sjh,Liu:2020evp,
Brito:2020lup,Baryakhtar:2020gao,Mai:2021yny,Xie:2022uvp,Bao:2022hew,Siemonsen:2022ivj,Cheng:2022jsw,
Zhou:2023sps,Jia:2023see,Bao:2023xna,Yang:2023vwm,Yang:2023aak,Dai:2023ewf,Dai:2023zcj,Chu:2024iie,Yu:2024mye,Xie:2025npy}, see also ref.~\cite{Brito:2015oca} for a comprehensive review). Systematic studies like ref.~\cite{Bernal:2022oha}, focusing on spin-$0$ DM candidates with masses above $10^3$~GeV, conclude that when the gravitational coupling is of order unity,\footnote{The gravitational coupling (or dimensionless mass coupling), usually denoted by $\alpha \equiv GM\mu/(\hbar c)$, relates the black hole mass $M$ and the particle mass $\mu$.} nearly extremal PBHs can boost final DM abundances by a factor of about $20$.

Even in the absence of PBHs, gravitational UV freeze-in provides another mechanism for DM production. In this process, SM particles at high energies annihilate through graviton exchange to produce pairs of DM particles, leading to an irreducible contribution to the DM abundance. This mechanism has been extensively studied~\cite{Cheek:2021cfe,Bernal:2020ili,Bernal:2018qlk,Tang:2017hvq,Garny:2017kha,Garny:2015sjg} and should be considered for a complete understanding of gravitational DM production.

In this work, we investigate the interplay among these gravitational mechanisms in DM production from PBHs that evaporate before Big Bang Nucleosynthesis (BBN) in the early universe, focusing particularly on how the combined effects of Hawking radiation, superradiant instabilities, and gravitational UV freeze-in can shape the final DM abundance for both scalar and vector DM. Specifically, we demonstrate that previous estimates of superradiant DM production~\cite{Bernal:2022oha} require revision when multiple-mode effects and gravitational wave (GW) emission from the superradiant cloud are properly accounted for. Our analysis reveals that if one considers
multiple superradiant modes rather than a single mode, the growth of each subsequent mode inevitably leads to the decay of the previous mode. Compared to the single-mode scenario, this multiple-mode evolution reduces rather than enhances the DM yields from superradiance. In addition,
the GW emission of the bound state substantially modifies the efficiency of the superradiant production mechanism. For scalar DM, while superradiance can still enhance production relative to pure Hawking radiation, the enhancement factor is smaller than previously estimated. For vector DM, our analysis indicates that if GW emission from the superradiant cloud proceeds sufficiently rapidly, it will deplete the cloud, resulting in a DM abundance lower than that produced solely by Hawking radiation.

The remainder of this paper is structured as follows. Section~\ref{sec:PBHs} introduces our assumptions on PBH formation and outlines the constraints imposed on initial PBH mass and abundance in our analysis. In section~\ref{sec:HRnSR}, we review the theoretical foundations of Hawking radiation and superradiance, with particular attention to multiple-mode dynamics and GW emission from superradiant clouds. We also provide analytic estimates that offer an overview of the evolution of the system before solving detailed equations. Section~\ref{sec:numevo} presents our results obtained by numerically solving the coupled evolution equations governing these processes. The contribution from gravitational UV freeze-in is examined in section~\ref{sec:UV}. Finally, section~\ref{sec:conclusion} summarizes our findings and discusses their implications.

Throughout this work, we employ natural units where $G=\hbar=c=k_{\mathrm{B}}=1$, and explicitly write out the Planck mass $M_{\mathrm{Pl}}\equiv\sqrt{{\hbar c}/{G}}$ when necessary for clarity.

\section{Primordial black holes in the early universe}\label{sec:PBHs}

PBHs can form through several mechanisms in the early universe. One mechanism involves the collapse of initial density inhomogeneities~\cite{Carr:1974nx,Carr:1975qj}. Alternative mechanisms include bubble collisions~\cite{Hawking:1982ga,Moss:1994iq,Khlopov:1999ys} and collapse of delayed false vacuum patches~\cite{Sato:1981bf,Maeda:1981gw,Sato:1981gv,Kodama:1981gu,Kodama:1982sf,Hsu:1990fg,Liu:2021svg,
Hashino:2021qoq,He:2022amv,Kawana:2022olo,Lewicki:2023ioy,Gehrman:2023esa,Gouttenoire:2023naa} during first-order phase transitions, and domain wall collapse during second-order phase transitions~\cite{Rubin:2000dq,Rubin:2001yw,Dokuchaev:2004kr}.\footnote{These domain walls can arise in several ways, such as spontaneous symmetry breaking~\cite{Rubin:2000dq} and quantum fluctuations during inflation~\cite{Linde:1990yj}.} In this work, rather than focusing on specific formation mechanisms, we adopt a monochromatic PBH mass spectrum to isolate and analyze PBH evolutionary effects independent of formation details. At an initial plasma temperature $T_\mathrm{ini}$, a typical estimate for the initial PBH mass scales with the horizon mass (the mass contained within the Hubble horizon)~\cite{Carr:1975qj,Carr:2020xqk},
\begin{equation}\label{eq:Mini}
M_\mathrm{ini}\equiv M_\mathrm{BH}\left(T_\mathrm{ini}\right) =\frac{4 \pi}{3} \gamma \frac{\rho_R\left(T_\mathrm{ini}\right)}{H^3\left(T_\mathrm{ini}\right)},
\end{equation}
with $\gamma = (1/3)^{3/2} \approx 0.2$ the assumed collapsing efficiency factor, $\rho_R$ the radiation energy density, and $H$ the Hubble parameter.

In our analysis, we require that PBHs fully evaporate before the onset of BBN, which leads us to impose an upper bound on the initial PBH mass~\cite{Carr:2020gox,Domenech:2020ssp}:
\begin{equation}\label{eq:BBN}
M_\mathrm{ini} \lesssim 10^9 \mathrm{~g}.
\end{equation}
Meanwhile, considering the PBHs that formed after reheating, a lower bound of $M_\mathrm{ini}$ can be imposed following CMB observations~\cite{Planck:2018jri}. Specifically, CMB observations constrain the tensor-to-scalar ratio, which in turn constrains the inflationary Hubble parameter $H_\mathrm{I}$ to be smaller than $\sim 10^{14} \mathrm{~GeV}$. Combined with eq.~\eqref{eq:Mini}, this upper bound on $H_\mathrm{I}$ translates to a lower bound on the initial PBH mass $M_\mathrm{ini}$:
\begin{equation}\label{eq:CMB}
    M_\mathrm{ini} \gtrsim 0.1 \mathrm{~g}.
\end{equation}

The initial abundance of PBH is characterized by the parameter $\beta$,
\begin{equation}
\beta \equiv \frac{\rho_{\mathrm{BH}}\left(T_\mathrm{ini}\right)}{\rho_R\left(T_\mathrm{ini}\right)},
\end{equation}
where $\rho_{\mathrm{BH}}$ and $\rho_R$ are energy densities of PBH and radiation, respectively. As the universe expands, the PBH energy density decreases as $a^{-3}$, with $a$ the scale factor, while the radiation energy density scales as $a^{-4}$. Given their different scaling behaviors, a PBH-radiation equality will be reached if the PBHs do not evaporate before this epoch. Then a PBH-dominated era will begin and last until PBH evaporation~\cite{Papanikolaou:2020qtd,Papanikolaou:2022chm,Domenech:2024wao}, which should be taken into account when considering the genesis of DM from the PBHs. The condition on $\beta$ for early matter domination to occur is~\cite{Domenech:2020ssp}
\begin{equation}\label{eq:PBHD}
\beta > \beta_c \approx 6.4 \times 10^{-10}\left(\frac{M_\mathrm{ini}}{10^4 \mathrm{~g}}\right)^{-1}.
\end{equation}
In addition, the energy density of the GWs at the time of BBN should satisfy $\Omega_{\mathrm{GW}, \mathrm{BBN}} \lesssim 0.05$~\cite{Caprini:2018mtu}, and this gives an upper bound of $\beta$, namely~\cite{Domenech:2020ssp}
\begin{equation}\label{eq:GW}
\beta \lesssim 1.1 \times 10^{-6}\left(\frac{M_\mathrm{ini}}{10^4 \mathrm{~g}}\right)^{-17 / 24}.
\end{equation}
One might also consider whether direct GW emission from PBH evaporation could contribute to the GW background, potentially alter eq.~\eqref{eq:GW}. However, as studied in ref.~\cite{Dong:2015yjs}, the spectral energy fraction today of GWs from PBH Hawking radiation reaches at most $\sim 10^{-7.5}$ for PBHs that evaporate before BBN.\footnote{The value $10^{-7.5}$ is the peak of the spectral energy fraction $\Omega_{\text{GW}}(f)$ at a frequency $f \approx 4 \times 10^{-14} \mathrm{~Hz}$. The integrated energy fraction does not greatly exceed the peak value in order of magnitude, so our argument here remains valid.} Using entropy conservation, this roughly translates to $\sim 4 \times 10^{-4}$, which is far below the BBN constraint $\Omega_{\mathrm{GW}, \mathrm{BBN}} \lesssim 0.05$. Therefore, the contribution of GWs from PBH Hawking radiation does not affect eq.~\eqref{eq:GW}.

The parameter space spanned by $M_\mathrm{ini}$ and $\beta$ delineates the conditions under which we analyze DM production. It encapsulates the distinct scenarios influencing the Hawking radiation of PBHs and the evolution of superradiance, which in turn affects the subsequent cosmological evolution and the resulting DM relic abundance.

\section{Hawking radiation and superradiance}\label{sec:HRnSR}

\subsection{Hawking radiation in Schwarzschild case}

A Schwarzschild BH of mass $M$ is characterized by an event horizon at $r_s\equiv2M$. Hawking's analysis~\cite{Hawking:1975vcx} revealed that BHs emit radiation from their event horizons, exhibiting thermal properties with a well-defined temperature. For a Schwarzschild BH, the horizon temperature is
\begin{equation}
T_{\mathrm{BH}}=\frac{1}{8 \pi M}.
\end{equation}
Hawking emission of particles with masses greater than $T_{\mathrm{BH}}$ is statistically suppressed, and only particles with smaller masses can be efficiently generated. For particle species $i$ with mass $\mu_i$, spin $s_i$, and internal degrees of freedom $g_i$, in the energy range $\left[E,E+\mathrm{d} E\right]$, the production rate from a Schwarzschild BH in Hawking radiation is~\cite{Cheek:2021odj}
\begin{equation}\label{eq:d2NidEdt}
\frac{\mathrm{d}^2 N_i}{\mathrm d E \mathrm d t}=\frac{g_i}{2 \pi} \frac{\Gamma_{s_i}\left(M_{\mathrm{BH}}, E \right)}{\exp \left(E / T_{\mathrm{BH}}\right)-(-1)^{2 s_i}},
\end{equation}
where $E=\sqrt{p^2+\mu_i^2}$ with $p$ the magnitude of $3$-momentum of particle. Here, the greybody factor $\Gamma_{s_i}$ is related to the absorption cross section $\sigma_{s_i}$ by $\Gamma_{s_i} \equiv \sigma_{s_i} p^2/\pi$~\cite{Page:1976df}. The greybody factor is present because the wave escaping from the BH propagates through the gravitational potential, resulting in a decreased intensity. The energy loss rate of the BH is given by the Page function $f\left(M\right)$~\cite{Page:1976df,Page:1976ki,Page:1977um}, namely
\begin{equation}\label{eq:Pagef}
f\left(M\right) \equiv-M^2 \frac{\mathrm{d} M}{\mathrm{~d} t} = M^2 \int_0^{+\infty} E \sum_i \frac{\mathrm{d}^2 N_i}{\mathrm d E \mathrm d t} \mathrm{~d} E.
\end{equation}

\subsection{Hawking radiation in Kerr case}

For a Kerr BH with mass $M$ and angular momentum $J \equiv M^2 a_*$, where $a_*$ is the dimensionless spin parameter,\footnote{Not to be confused with the scale factor $a$.} the metric in Boyer–Lindquist coordinates reads
\begin{equation}
\begin{aligned}
\mathrm{d} s^2= & -\mathrm{d} t^2+\frac{2 M r}{\Sigma}\left(\mathrm{d} t- M a_{*} \sin ^2 \theta \mathrm{d} \varphi\right)^2 \\
& +\frac{\Sigma}{\Delta} \mathrm{d} r^2+\Sigma \mathrm{d} \theta^2+\left(r^2+ M^2 a_{*}^2\right) \sin \theta^2 \mathrm{d} \varphi^2,
\end{aligned}
\end{equation}
where $\Sigma \equiv r^2+ M^2 a_{*}^2 \cos ^2 \theta$ and $\Delta \equiv r^2-2 M r+ M^2 a_{*}^2$.  A Kerr BH has an outer event horizon and an inner Cauchy horizon located at $r=r_+$ and $r=r_-$, which are given by
\begin{equation}
r_{ \pm}=M\left(1 \pm \sqrt{1-a_{*}^2}\right).
\end{equation}
The angular velocity of the outer horizon is $\Omega_\mathrm{H} = a_*/(2r_+)$. For the Kerr BH, the horizon temperature is
\begin{equation}
T_{\mathrm{BH}}=\frac{\sqrt{1-a_{*}^2}}{4 \pi r_+} .
\end{equation}
The particle production rate is
\begin{equation}
\frac{\mathrm{d}^2 N_i}{\mathrm{d} E \mathrm{d} t}=\frac{g_i}{2 \pi} \sum_{l=s} \sum_{m=-l}^l \frac{\mathrm{d}^2 N_{i, l m}}{\mathrm{d} E \mathrm{d} t}
\end{equation}
with
\begin{equation}
\frac{\mathrm{d}^2 N_{i, l m}}{\mathrm{d} E \mathrm{d} t}=\frac{\Gamma_{s_i}^{l m}\left(M_{\mathrm{BH}}, E, a_*\right)}{\exp \left[\left(E-m \Omega_\mathrm H\right) / T_{\mathrm{BH}}\right]-(-1)^{2 s_i}},
\end{equation}
where $\Gamma_{s_i}^{l m}$ is the corresponding greybody factor. The definition of the Page function $f\left(M, a_*\right)$ in the Kerr case has the same formula as in the Schwarzschild case. Another Page function $g\left(M, a_*\right)$ gives the angular momentum loss rate,
\begin{equation}
\begin{aligned}
g\left(M, a_*\right) & \equiv-\frac{M}{a_*} \frac{\mathrm{d} J}{\mathrm{~d} t} = -\frac{M}{a_*} \int_0^{+\infty} \sum_i g_i \sum_{l, m} m \frac{\mathrm{d}^2 N_{i, l m}}{\mathrm{~d} E \mathrm{~d} t}\mathrm{~d} E.
\end{aligned}
\end{equation}
The equations for the evolution of $M$ and $a_*$ of a Kerr BH under Hawking radiation are then
\begin{align}
\frac{\mathrm{d} M}{\mathrm{~d} t}&=-\frac{f(M, a_*)}{M^2},  \\
\frac{\mathrm{d} a_*}{\mathrm{~d} t} & =a_* \frac{2 f\left(M, a_*\right)-g\left(M, a_*\right)}{M^3} .
\end{align}

We assume all PBHs are born with a universal
large dimensionless spin parameter $a_{*\mathrm{ini}}$, which is unlikely with the usual PBH formation mechanism based on
a horizon-size collapse of density perturbations (see relevant discussions and extensive studies on PBH spins in refs.~\cite{Chiba:2017rvs,Mirbabayi:2019uph,He:2019cdb,Flores:2021tmc,Chongchitnan:2021ehn,
Eroshenko:2021sez,Garcia-Bellido:2020pwq,Harada:2020pzb,Banerjee:2024nkv}).
\footnote{For example, Harada et al.\cite{Harada:2020pzb} investigated the spins of PBHs formed during the radiation-dominated era. They found that the initial spin is of order $10^{-3}$ or even smaller for PBHs with masses comparable to the horizon mass ($M \sim M_\mathrm{H}$), supporting the statement that such PBHs typically possess small spins.} Nevertheless,
as discussed in section~\ref{sec:intro}, there exist alternative mechanisms which may give rise to large initial
spins for PBHs. In this work we adopt a phenomenological and
practical approach with a simple universal $a_{*\mathrm{ini}}$ as in ref.~\cite{Bernal:2022oha}, being agnostic about the specific
mechanism behind the scene. The consideration of
specific mechanisms and their impact on the results will be left for future work.

\subsection{Estimations for Hawking radiation}

To fully determine the BH evolution under Hawking radiation, it is essential to include the greybody factors for all particles. This task is non-trivial, especially when considering the effects of angular momenta in Kerr BHs and vector DM particles. In this work, we employ the public package \texttt{BlackHawk}~\cite{Arbey:2019mbc} to evaluate the relative quantities numerically. Although fully including greybody factors requires complex numerical calculations, we can still obtain analytical insights by considering the simple case of scalar DM production from Schwarzschild PBHs.

In the Schwarzschild case, the absorption cross-section $\sigma_{s_i}$ for massless particles approaches the geometrical-optics limit $\sigma_{s_i}\rightarrow27\pi M^2$ when $E/T_\mathrm{BH}~\gg~1$. In the geometrical-optics limit,\footnote{Here we employ the geometrical-optics limit only for estimation purposes. In the subsequent section, the complete treatment, including numerical evaluations of the greybody factors, is employed.} the Page function $f(M)$ is~\cite{Baldes:2020nuv,Cheek:2021odj}
\begin{equation}
f(M)=\frac{27}{4} \frac{g_{*}\left(T_\mathrm{BH}\right)}{30720 \pi},
\end{equation}
where $g_{*}\left(T_\mathrm{BH}\right)$ is the effective number of relativistic degrees of freedom at $T_\mathrm{BH}$, which does not vary significantly in this work.\footnote{In the range of initial PBH mass that we consider, $T_\mathrm{BH}$ is greater than the mass of the top quark $\mu_\mathrm{top} \approx 173 \mathrm{~GeV}$.} To estimate the PBH lifetime, we can roughly treat $f(M)$ as a constant, so eq.~\eqref{eq:Pagef} yields an approximate evolution for the BH mass:
\begin{equation}
M(t)\approx M_\mathrm{ini}\left(1-\frac{t-t_\mathrm{ini}}{\tau_\mathrm{BH}}\right)^{1 / 3},
\end{equation}
where $t_\mathrm{ini}$ is the time at PBH formation, and
\begin{equation}
    \tau_\mathrm{BH} \equiv \frac{M_\mathrm{ini}^3}{3f\left(M_\mathrm{ini}\right)}
\end{equation}
is the approximate PBH lifetime. We may choose $g_{*,\text{SM}}(T_\mathrm{BH})=106.75$ and also include gravitons, which have degrees of freedom $g_\text{grav}=2$. The timescale for PBH life is then
\begin{equation}\label{eq:PBH_life}
\tau_\mathrm{BH} = \frac{1}{3} \left(\frac{27}{4} \frac{106.75+2}{30720 \pi}\right)^{-1} M_\mathrm{ini}^3
\approx  2.3 \times 10^{-19} \mathrm{~s}\left(\frac{M_\mathrm{ini}}{1 \mathrm{~kg}}\right)^3.
\end{equation}

Note that eq.~\eqref{eq:PBH_life} does not take into account the Hawking emission of DM particles. By assuming DM particle mass $\mu \rightarrow 0$ when integrating eq.~\eqref{eq:d2NidEdt}, the number of scalar DM particles from Hawking radiation of a Schwarzschild BH reads
\begin{equation}\label{eq:NhrDM}
\begin{aligned}
N_\mathrm{DM}^{\mathrm{hr}} & =\int_{t_\mathrm{ini}}^{\tau_{\mathrm{BH}}} \frac{\mathrm{~d} N_\mathrm{DM}^{\mathrm{hr}}}{\mathrm{~d} t} \mathrm{~d} t =\int_{M_\mathrm{ini}}^0\left(\frac{\mathrm{~d} N_\mathrm{DM}^\mathrm{hr}}{\mathrm{~d} t}\right)\left(\frac{\mathrm{~d} M}{\mathrm{~d} t}\right)^{-1} \mathrm{~d} M \\
& =\frac{120 \zeta(3)}{\pi^3} \frac{g_\mathrm{DM}}{g_{*}\left(T_{\mathrm{BH}}\right)} \frac{M_\mathrm{ini}^2}{M_{\mathrm{Pl}}^2} \approx 8.9 \times 10^{13} \left(\frac{M_\mathrm{ini}}{1 \mathrm{~kg}}\right)^2.
\end{aligned}
\end{equation}

\subsection{Growth of a superradiant mode}\label{subsec:growthsm}

A BH with an ergoregion can spontaneously trigger superradiant instabilities, wherein bosonic modes with Compton wavelengths comparable to the BH's gravitational radius form quasibound states with large occupation numbers. The resulting configuration---often called a superradiant cloud---is analogous to the electron cloud surrounding an atomic nucleus,\footnote{The analogy is not complete, since the electrons obey the Pauli exclusion principle, while the superradiant cloud allows for
large occupation numbers.} prompting the term ``gravitational atom".\footnote{It is important to distinguish between superradiant amplification associated with
free wave scattering from rotating objects, and superradiant instability, the feedback mechanism responsible for the exponential growth of quasibound states. We focus here on the latter phenomenon.}

In the Kerr geometry, a superradiant mode is characterized by the spacetime
dependence of the field $e^{-i\omega t+im\varphi}R(r)S(\theta)$, with $t,r,\theta,\varphi$
being the corresponding Boyer-Lindquist coordinates. For quasibound states the angular
frequency $\omega$ is discrete and complex, with a small positive imaginary part
characterizing the superradiant instability. To characterize the discreteness
of $\omega$ we introduce an overtone number $n$ so that the angular frequency can
be labeled as $\omega_n$, with $n=1$ denoting the lowest-frequency state. The magnetic number $m$ is
associated with the Killing vector $k_z\equiv -i\frac{\partial}{\partial\varphi}$.
In the general case there are two additional angular momentum numbers $l$ and $j$, which can
be associated with the orbital angular momentum and the total angular momentum in the
flat-space limit. Therefore for vector DM a quasibound mode can be represented by
$\left|nljm\right\rangle$ (called a vector mode) while for scalar DM a quasibound mode
can be represented by $\left|nlm\right\rangle$ (called a scalar mode) since the total angular momentum and the orbital angular momentum are the same for a scalar field.

The superradiant instability occurs as long as the following \emph{superradiance condition}
is satisfied
\begin{equation}\label{eq:src}
    m \Omega_\mathrm H-\omega_R>0.
\end{equation}
Hereafter $\omega_R$ denotes the real part of the angular frequency of the mode (see appendix~\ref{ap:sr_modes}). $\Omega_\mathrm H \equiv a_*/(2r_+)$ is the angular velocity of the outer horizon of the Kerr BH. As superradiance occurs, the energy and angular momentum of the BH are extracted to the superradiant mode, leading to the exponential growth of the occupation number. When $m \Omega_\mathrm H-\omega_R=0$, the mode becomes saturated and stops growing. When $m \Omega_\mathrm H-\omega_R<0$, the corresponding mode decays
rather than grow.

In this work, we consider superradiant DM particles that neither exhibit self-interactions nor couple non-gravitationally with SM particles. For a scalar mode $\left|nlm\right\rangle$, the superradiant instability rate $\Gamma^\mathrm{sr}$ is roughly (see appendix~\ref{ap:sr_modes} for complete analytic expressions)~\cite{Baumann:2019eav,Brito:2015oca,Bernal:2022oha}
\begin{equation}\label{eq:GammasrS}
    \Gamma_{nlm}^\mathrm{sr} \approx 2\mu C_{nl}\left(a_{*}m-2\mu r_+\right)\left(M\mu\right)^{4l+4},
\end{equation}
For a vector mode $\left|nljm\right\rangle$ the instability rate reads
\begin{equation}\label{eq:GammasrV}
    \Gamma_{nljm}^\mathrm{sr} \approx 2\mu C_{nlj}\left(a_{*}m-2\mu r_+\right)\left(M\mu\right)^{2j+2l+4}.
\end{equation}
We note that $n$ is analogous to its counterpart in quantum mechanics, labeling the energy levels, i.e.,
\begin{equation}
    \omega_R \approx \mu\left(1-\frac{M^2\mu^2}{2 n^2}\right).
\end{equation}
Note that for the analytic approximations to hold, the dimensionless mass coupling $M\mu$ must be significantly smaller than $1$, and the above equation applies to both the scalar and the vector DM cases to the displayed order. Higher-order terms in $M\mu$ introduce corrections to the energy levels, which we account for when computing the superradiant evolution. Details on these corrections are also provided in appendix~\ref{ap:sr_modes}.

The coefficients $C_{nl}$ and $C_{nlj}$ in eqs.~\eqref{eq:GammasrS} and \eqref{eq:GammasrV} are~\cite{Baumann:2019eav}
\begin{equation}
C_{n l} \equiv \frac{2^{4 l+1}(n+l)!}{n^{2 l+4}(n-l-1)!}\left[\frac{l!}{(2 l)!(2 l+1)!}\right]^2,
\end{equation}
and
\begin{equation}
\begin{aligned}
C_{n l j} \equiv &\frac{2^{2 l+2 j+1}(n+l)!}{n^{2 l+4}(n-l-1)!}\left[\frac{(l)!}{(l+j)!(l+j+1)!}\right]^2\\
&\times\left[1+\frac{2(1+l-j)(1-l+j)}{l+j}\right]^2.
\end{aligned}
\end{equation}
Then the instability timescale of a mode can be defined as $\tau^\mathrm{sr}\equiv1/\Gamma^\mathrm{sr}$.

If the instability timescale of a superradiant mode is significantly shorter than the BH lifetime, superradiance for this mode can occur, and a maximum occupation number will be reached. For a mode with magnetic number $m$, and a BH with mass $M$ and some initial spin $a_{*i}$, the superradiance condition \eqref{eq:src} holds if $a_{*i}$ exceeds the critical spin $a_{*c}$. Setting $m \Omega_\mathrm H-\omega_R=0$, we can approximate $a_{*c}$ as
\begin{equation}
\label{eq:a*c}
a_{* c} \approx \frac{4 m M \mu}{m^2+4 M^2 \mu^2}\approx\frac{0.4}{m}\frac{M \mu}{0.1}.
\end{equation}
where the second approximation holds under the assumption that $M\mu \ll 1$. The maximum occupation number for this mode is then given approximately by~\cite{Baryakhtar:2017ngi}
\begin{equation}\label{eq:Nsr_max}
N_{\max}^{\mathrm{sr}} \approx M^2 \frac{a_{* i}-a_{* c}}{m} \approx 2.11 \times 10^{15}\left(\frac{M}{1 \mathrm{~kg}}\right)^2\frac{a_{* i}-a_{* c}}{m},
\end{equation}
and is the same for both scalar and vector modes.

The fastest-growing modes for the scalar and vector fields are $\left|nlm\right\rangle = \left|211\right\rangle$ and $\left|nljm\right\rangle = \left|1011\right\rangle$, respectively.  Their growth rates are (see also refs.~\cite{Baryakhtar:2017ngi, Brito:2015oca})
\begin{equation}
    \Gamma^\mathrm{sr}_\mathrm{211} \approx \frac{1}{24}\mu \left(a_{*}-2\mu r_+\right)\left(M\mu\right)^{8}
\end{equation}
and
\begin{equation}
    \Gamma^\mathrm{sr}_\mathrm{1011} \approx 4\mu \left(a_{*}-2\mu r_+\right)\left(M\mu\right)^{6}.
\end{equation}
The corresponding instability timescales are (see also ref.~\cite{Brito:2017zvb})
\begin{equation}\label{eq:211_time}
\tau^\mathrm{sr}_{211} \approx 5.94 \times 10^{-26} \mathrm{~s}\left(\frac{M}{1 \mathrm{~kg}}\right)\left(\frac{0.1}{M\mu}\right)^9 \frac{1}{a_*}
\end{equation}
and
\begin{equation}\label{eq:1011_time}
\tau^\mathrm{sr}_{1011} \approx 6.19 \times 10^{-30} \mathrm{~s}\left(\frac{M}{1 \mathrm{~kg}}\right)\left(\frac{0.1}{M\mu}\right)^7 \frac{1}{a_*}.
\end{equation}
where $1/(a_{*i}-a_{*c})$ is approximated by $1/a_*$ for simplicity.

Note that multiple superradiant modes can coexist. In this work, for each of the scalar and vector cases, we consider the coexistence of two modes. For the scalar case, they are $\left|211\right\rangle$ and $\left|322\right\rangle$. The growth rate and timescale for $\left|322\right\rangle$ mode are
\begin{align}
    \Gamma^\mathrm{sr}_{322}& \approx \frac{8}{885735} \mu \left(2a_{*}-2\mu r_+\right)\left(M\mu\right)^{12},\\
    \tau^\mathrm{sr}_{322}&\approx 1.37 \times 10^{-18} \mathrm{~s}\left(\frac{M}{1 \mathrm{~kg}}\right)\left(\frac{0.1}{M\mu}\right)^{13} \frac{1}{a_*}.\label{eq:322_time}
\end{align}
For the vector case, the two modes are $\left|1011\right\rangle$ and $\left|2122\right\rangle$, and the growth rate and timescale for $\left|2122\right\rangle$ are
\begin{align}
    \Gamma^\mathrm{sr}_{2122}& \approx \frac{1}{864} \mu \left(2a_{*}-2\mu r_+\right)\left(M\mu\right)^{10},\\
    \tau^\mathrm{sr}_{2122}&\approx 1.07 \times 10^{-22} \mathrm{~s}\left(\frac{M}{1 \mathrm{~kg}}\right)\left(\frac{0.1}{M\mu}\right)^{11} \frac{1}{a_*}.\label{eq:2122_time}
\end{align}
Several considerations guided our selection of the $\left|322\right\rangle$ and $\left|2122\right\rangle$ modes as the second modes for the scalar and vector cases, respectively. For a detailed discussion, we refer readers to appendix~\ref{ap:sr_modes}.

If Hawking radiation and superradiant population were the only two factors determining the number of DM particles, we could compare eq.~\eqref{eq:Nsr_max} with eq.~\eqref{eq:NhrDM} to conclude that superradiance could enhance the generation of DM from PBHs by about an order of magnitude. This enhancement would reduce the initial abundance of PBHs required to produce DM. However, the situation is complicated by the existence of multiple modes, and also counterbalanced by another effect of superradiance: GW emission from the superradiant cloud.

\subsection{GW emission of the superradiant cloud}

The superradiant cloud emits monochromatic GWs with angular frequency $\omega_\mathrm{GW} \approx 2 \mu$. This process can be interpreted as the annihilation of two superradiant particles in the cloud to produce one graviton~\cite{Arvanitaki:2010sy} and thus decreases the occupation number.\footnote{Additional GW emission channels such as energy-level transitions between two modes and
annihilation of two different modes have been neglected in this work; see appendix~\ref{ap:trans_GWs} for justification.} For the superradiant modes we consider, the GW powers are~\cite{Guo:2022mpr, Guo:2024dqd}
\begin{align}
P^{\mathrm{GW}}_{211}&\approx\frac{484+9 \pi^2}{23040}\left(\frac{M^\mathrm{sr}_{211}}{M}\right)^2(M \mu)^{14}, \\
P^{\mathrm{GW}}_{322}&\approx\frac{1024+49 \pi^2}{5423886846}\left(\frac{M^\mathrm{sr}_{322}}{M}\right)^2(M \mu)^{18},\\
P^{\mathrm{GW}}_{1011} &\approx \frac{8\left(64+9 \pi^2\right)}{45}\left(\frac{M^\mathrm{sr}_{1011}}{M}\right)^2(M \mu)^{10},\\
P^{\mathrm{GW}}_{2122} &\approx \frac{4096+1225 \pi^2}{1433600}\left(\frac{M^\mathrm{sr}_{2122}}{M}\right)^2(M \mu)^{14},
\end{align}
where $M^\mathrm{sr}_{nlm}$ or $M^\mathrm{sr}_{nljm}$ denotes the cloud mass of the corresponding mode. These expressions are valid when $M\mu$ is small, but they also correctly scale for moderately large $M\mu$~\cite{Brito:2014wla, Baryakhtar:2017ngi, Brito:2017zvb,Yoshino:2013ofa,Siemonsen:2019ebd,Brito:2015oca}. The GW emission is more powerful for the vector case because the emitted GW flux scales as $\left({M^\mathrm{sr}}/{M}\right)^2(M\mu)^{4 l+10}$ and the sequence of superradiant vector modes starts from $l=0$ instead of $l=1$ in the scalar case.

The GW emission persists until the cloud is depleted. Setting $t=0$ as the moment when superradiance saturates, and assuming constant BH mass and spin, the mass of the superradiant cloud decreases exclusively due to its GW emission, namely~\cite{Brito:2017zvb}
\begin{equation}
    \frac{\mathrm{d} M^{\mathrm{sr}}}{\mathrm{~d} t}=-P^{\mathrm{GW}}.
\end{equation}
Consequently, a typical GW emission timescale $\tau^{\mathrm{GW}}$ can be defined based on the cloud mass decay pattern:
\begin{equation}
M^\mathrm{sr}(t)=\frac{M^\mathrm{sr}_{\max}}{1+t / \tau^{\mathrm{GW}}},
\end{equation}
We can then read off $\tau^{\mathrm{GW}}$ for our considered superradiant modes:
\begin{align}
\tau^{\mathrm{GW}}_{211} &\approx 1.0 \times 10^{-19}\mathrm{~s} \left( \frac{M}{1\mathrm{~kg}} \right)\left( \frac{0.1}{M\mu} \right)^{15}\frac{1}{a_*}, \label{eq:211_gw_time} \\
\tau^{\mathrm{GW}}_{322} &\approx 1.8 \times 10^{-10}\mathrm{~s} \left( \frac{M}{1\mathrm{~kg}} \right)\left( \frac{0.1}{M\mu} \right)^{19}\frac{1}{a_*},\\
\tau^{\mathrm{GW}}_{1011} &\approx 9.1 \times 10^{-27}\mathrm{~s} \left( \frac{M}{1\mathrm{~kg}} \right)\left( \frac{0.1}{M\mu} \right)^{11}\frac{1}{a_*},\\
\tau^{\mathrm{GW}}_{2122} &\approx 4.4 \times 10^{-19}\mathrm{~s} \left( \frac{M}{1\mathrm{~kg}} \right)\left( \frac{0.1}{M\mu} \right)^{15}\frac{1}{a_*}. \label{eq:2122_gw_time}
\end{align}
Now that we have all the relevant timescales, we can proceed with the estimation of the evolution. By comparing the PBH lifetime \eqref{eq:PBH_life} with the superradiance timescales \eqref{eq:211_time}, \eqref{eq:1011_time}, \eqref{eq:322_time}, and \eqref{eq:2122_time}, as well as the GW timescales \eqref{eq:211_gw_time} through \eqref{eq:2122_gw_time}, we can make the following general observation: For the fastest superradiant mode, its superradiance timescale is probably
significantly shorter than the GW timescale, so that a maximum occupation number \eqref{eq:Nsr_max} for a superradiant mode can be attained. Once a superradiant mode reaches saturation, its associated cloud continuously dissipates due to GW emission with a timescale which can exceed or fall short of
the PBH lifetime. Moreover, the fastest superradiant mode may be depleted due to the growth of the secondary mode, depending on the BH and DM mass parameters. The secondary mode evolves further under the influence of its own GW emission until PBH evaporation.

Although superradiance of the fastest mode can generate more DM particles than Hawking radiation, the situation is complicated by the coexistence of multiple modes and their GW emission. Therefore, it is essential to conduct a detailed investigation into the evolution of the system over time, which will be the focus of the next section.

\section{Evolution under Hawking radiation and superradiance}\label{sec:numevo}

The evolution of a single PBH governed by Hawking radiation and superradiance is determined by a set of differential equations:
\begin{align}
\frac{\mathrm d N_i^{\mathrm{sr}}}{\mathrm d t}&=\Gamma_i^{\mathrm{sr}} N_i^{\mathrm{sr}}-\frac{P_i^\mathrm{G W}}{\mu}, \label{eq:evo41} \\
\frac{\mathrm d M}{\mathrm d t}&=-\frac{f\left(M, a_*\right)}{M^2}- \sum_i \mu \Gamma_i^{\mathrm{sr}} N_i^{\mathrm{sr}}, \label{eq:evo42} \\
\frac{\mathrm{d} a_*}{\mathrm{~d} t}&=-a_* \frac{g\left(M, a_*\right)-2 f\left(M, a_*\right)}{M^3}- \left(\sum_i \frac{m_i}{M^2} \Gamma_i^{\mathrm{sr}} N_i^{\mathrm{sr}}-\frac{2a_*}{M}\sum_i \mu \Gamma_i^{\mathrm{sr}} N_i^{\mathrm{sr}}\right), \label{eq:evo43}
\end{align}
where $i$ denotes the $i$-th superradiant mode that we consider, so for example $m_i$ is the magnetic number of the $i$-th supperadiant mode, etc. During the growing phase of superradiance, the occupation number $N_i^{\mathrm{sr}}$ increases at the rate $\Gamma_i^\mathrm{sr}$, but is simultaneously decreased by GW emission at the power $P_i^\mathrm{GW}$. The PBH loses its mass and angular momentum in the presence of both Hawking radiation and superradiance.

We note that a superradiant mode may contribute significantly to the DM yield only if its instability timescale is somewhat shorter than the PBH lifetime. Using the analytic estimates for $\tau_\mathrm{BH},\tau^\mathrm{sr}_{211},\tau^\mathrm{sr}_{322},
\tau^\mathrm{sr}_{1011},\tau^\mathrm{sr}_{2122}$ derived in subsection~\ref{subsec:growthsm},
we may infer that for this requirement is usually satisfied for the first mode for $M\mu\sim\mathcal{O}(0.1)$\footnote{$M\mu\gg(0.1)$ or $M\mu\ll(0.1)$ will lead to highly suppressed
superradiant instability rate~\cite{Arvanitaki:2010sy}.}, while for the second mode its instability timescale can be
shorter, longer or comparable to the PBH lifetime depending on the parameters. Moreover,
the superradiance condition eq.~\eqref{eq:src} which governs the ending of the instability
involves quantities that evolve dynamically. Facing these complications, we have numerically implemented eqs.~\eqref{eq:evo41}\eqref{eq:evo42}\eqref{eq:evo43} in our calculation, taking into account the Hawking radiation, superradiance, together with their interplay, in a fully dynamical manner. Especially, the superradiance condition eq.~\eqref{eq:src} is evaluated dynamically, taking into account time-dependent evolution of PBH masses and angular momenta. This dynamical implementation is adopted for all points of parameter space in the parameter scan below.

The validity of our computational framework relies on two key assumptions. First, the quasi-adiabatic approximation should hold, which requires that the evolution of the scalar cloud under superradiant growth and GW emission is sufficiently slow compared to the dynamical timescale of the PBH. Second, the backreaction of the superradiant cloud on the spacetime metric can be neglected. This assumption holds as long as the mass of the superradiant cloud remains small compared to that of the PBH.

In practice, these two assumptions break down once the PBH mass is sufficiently reduced by Hawking radiation. Therefore, we manually terminate the computation at very small PBH masses\footnote{The breakdown points associated with the adiabatic and the backreaction considerations can be inferred from the analysis in ref.~\cite{March-Russell:2022zll}.}. We expect this cutoff to have a minimal impact on our analysis, because Hawking radiation becomes explosive at late times, and the change of the key quantities, such as the DM number density, remains relatively small during this very short period of time.


Furthermore, the superradiant DM cloud is theoretically modeled as a classical field surrounding the PBH during the superradiance process, which naturally raises questions about its fate following complete PBH evaporation. Analyses in the literature~\cite{Calza:2023rjt, March-Russell:2022zll} suggest that due to quenching-like effects, the DM cloud may evolve into a self-gravitating configuration before the PBH fully evaporates, forming a microscopic boson star afterwards. Nonetheless, for clarity and convenience in subsequent discussions, we will continue to employ terminology such as ``DM particle".

The early universe consists of three distinct components of energy density: PBHs, SM particles considered as radiation, and DM. Therefore, the Friedmann equation that accounts for these three contributions is expressed as\footnote{We neglect GW emission as part of the radiation component. As demonstrated in the subsequent discussion, in the case of PBH domination, the superradiant effect is negligible; conversely, in the absence of PBH domination, the amount of GW emission from the superradiant cloud is insignificant compared to the whole radiation component.}
\begin{equation}
\frac{3 H^2}{8 \pi}=\rho_{\mathrm{PBH}}+\rho_{\mathrm{SM}}+\rho_{\mathrm{DM}},
\end{equation}
with
\begin{align}
\frac{\mathrm{d}\rho_{\mathrm{PBH}}}{\mathrm{d} t}+3 H \rho_{\mathrm{PBH}}&=\frac{\rho_{\mathrm{PBH}}}{M} \frac{\mathrm{d} M}{\mathrm{d} t} ,\\
\frac{\mathrm{d}\rho_{\mathrm{SM}}}{\mathrm d t}+4 H \rho_{\mathrm{SM}}&=-\frac{\rho_{\mathrm{PBH}}}{M}\left.\frac{\mathrm{d}M}{\mathrm d t}\right|_{\mathrm{SM}},\label{eq:SMevo} \\
\frac{\mathrm{d} \rho_{\mathrm{DM}}}{\mathrm{~d} t}+3 H \rho_{\mathrm{DM}}&=\frac{\rho_{\mathrm{PBH}}}{M}\mu\left(\frac{\mathrm{~d} N^{\mathrm{hr}}}{\mathrm{~d} t}+\sum_i \frac{\mathrm{~d} N_i^{\mathrm{sr}}}{\mathrm{~d} t}\right).
\end{align}
Here $\left.\frac{\mathrm{d}M}{\mathrm d t}\right|_{\mathrm{SM}}$ represents the rate of
BH mass loss due to Hawking emission of SM particles. We consider DM particles that are
heavier than the SM top quark so that they behave non-relativistically after being
produced from Hawking radiation or superradiance. Note that eq.~\eqref{eq:SMevo} accounts for
the heating of the primordial plasma due to the evaporation of PBHs.

We can numerically evaluate these equations in the time interval between PBH formation $t_\mathrm{ini}$ and the end of PBH evaporation $t_\mathrm{ev}$ to obtain the evolution during this period. After the PBHs have fully evaporated, we assume entropy conservation and calculate the DM relic abundance using
\begin{equation}
\Omega_{\mathrm{DM}}h^2=\frac{h^2}{\rho_{c,0}} \frac{g_{\star s}\left(T_0\right) T_0^3}{g_{\star s}\left(T_{\mathrm{ev}}\right) T_{\mathrm{ev}}^3} \rho_{\mathrm{DM,ev}},
\end{equation}
where $h$ is defined by $H_0\equiv100\,h\,\mathrm{km}\,\mathrm{s}^{-1}\,\mathrm{Mpc}^{-1}$ with $H_0$ being the Hubble constant. $\rho_{c,0}\approx1.05 \times 10^{-5} h^2\mathrm{~GeV}\cdot\mathrm{cm}^{-3}$ and $T_0=2.73\,\mathrm{K}$ are respectively the critical density and the CMB temperature of today's universe, $\rho_{\mathrm{DM,ev}}$ and $T_\mathrm{ev}$ are resepctively the energy density of DM and the temperature of radiation at the end of PBH evaporation, and $g_{\star s}$ is the effective number of degrees of freedom in entropy which is a function of temperature under consideration.

\begin{figure}[t]
    \centering
    \includegraphics[width=0.32\textwidth]{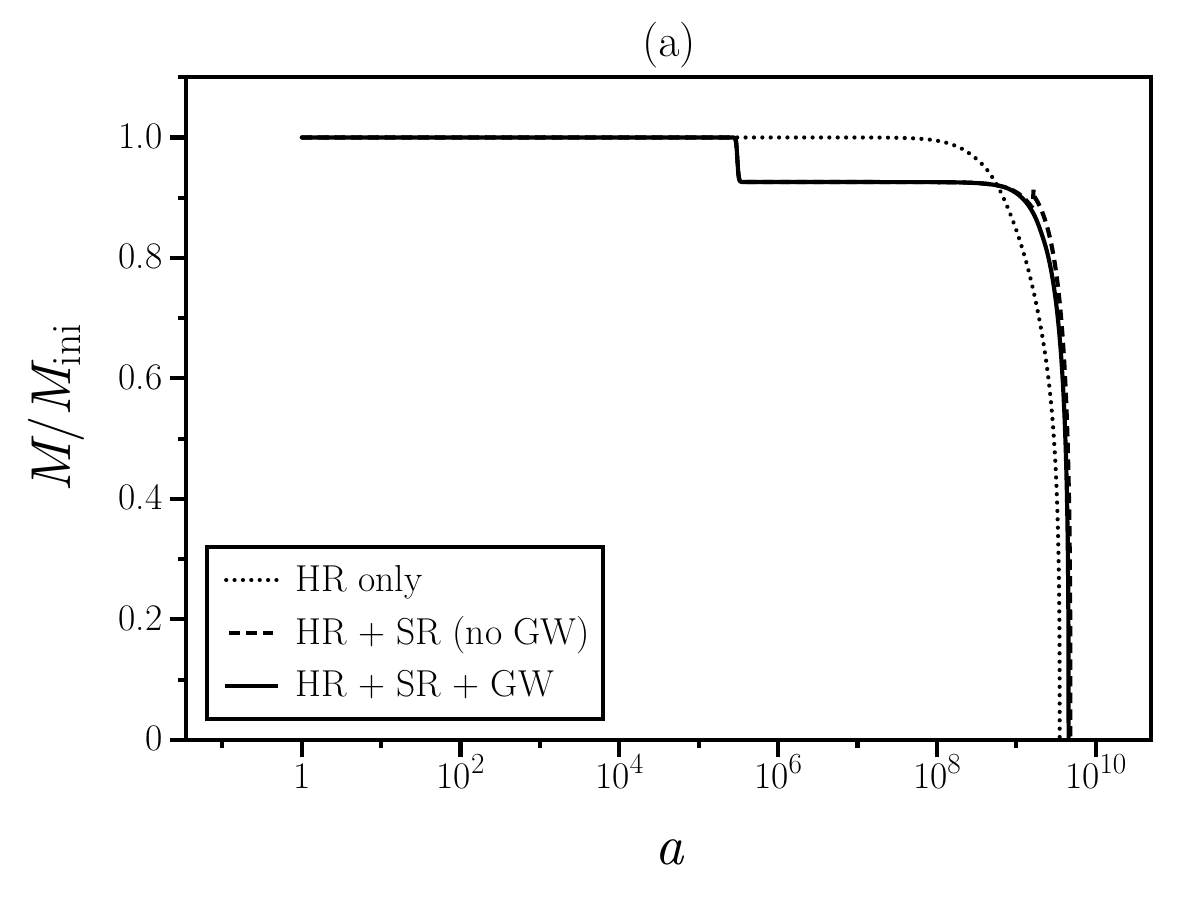}
    \includegraphics[width=0.32\textwidth]{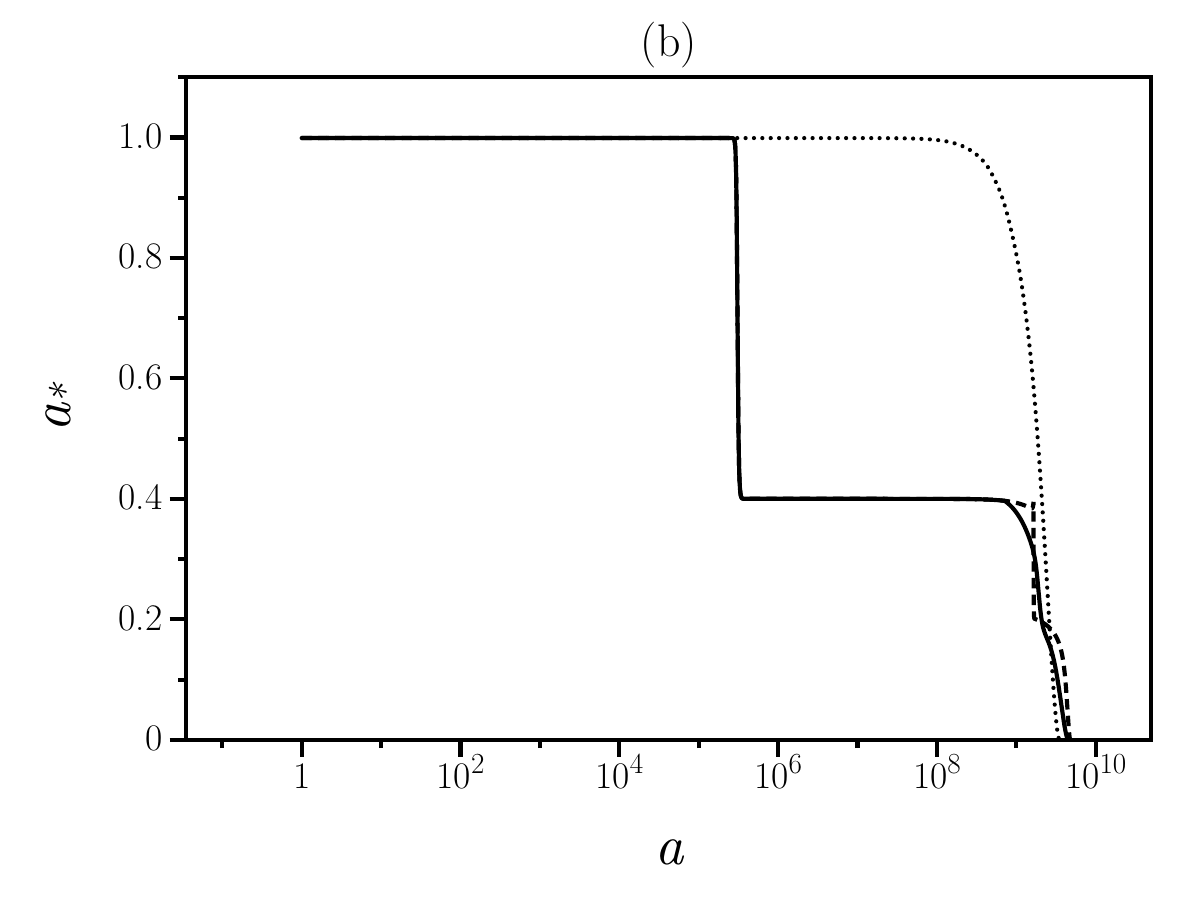}
    \includegraphics[width=0.32\textwidth]{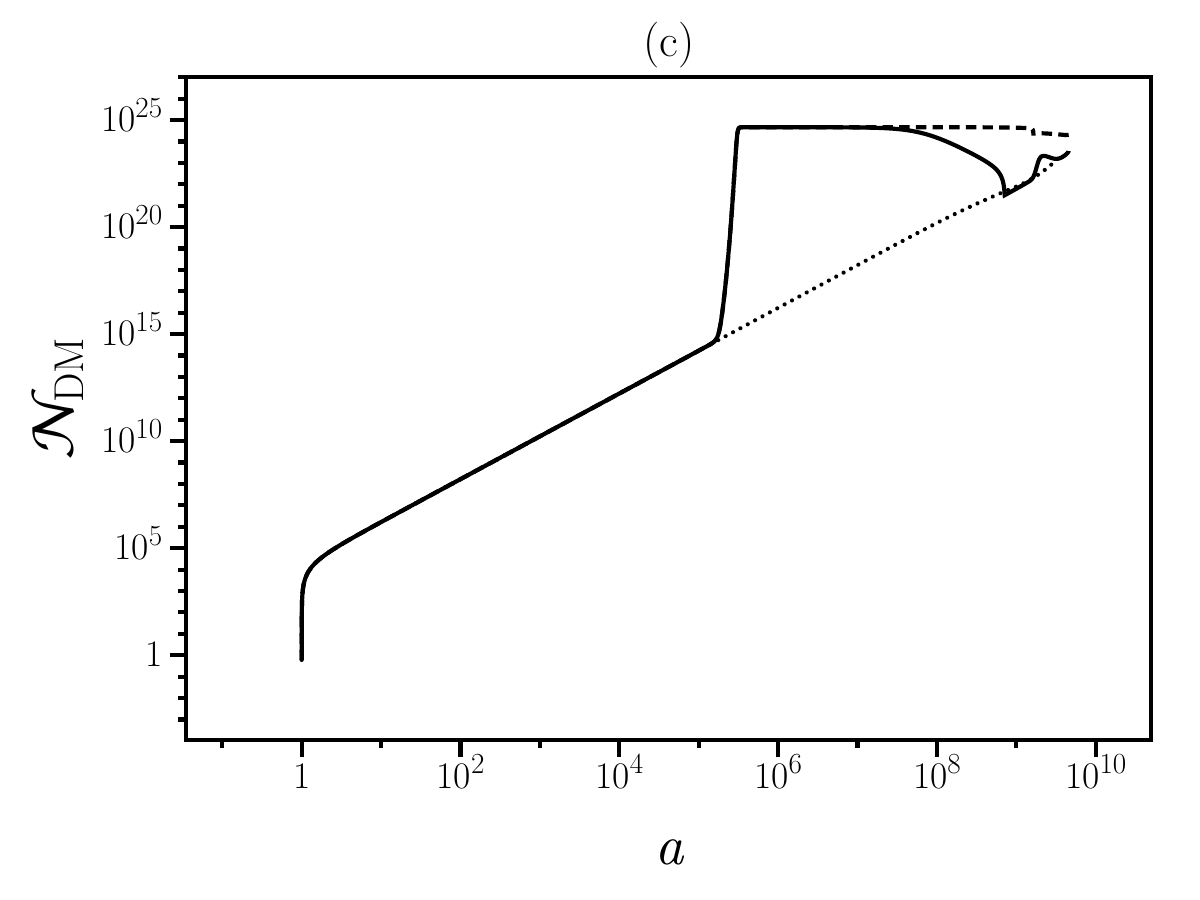}
    \includegraphics[width=0.32\linewidth]{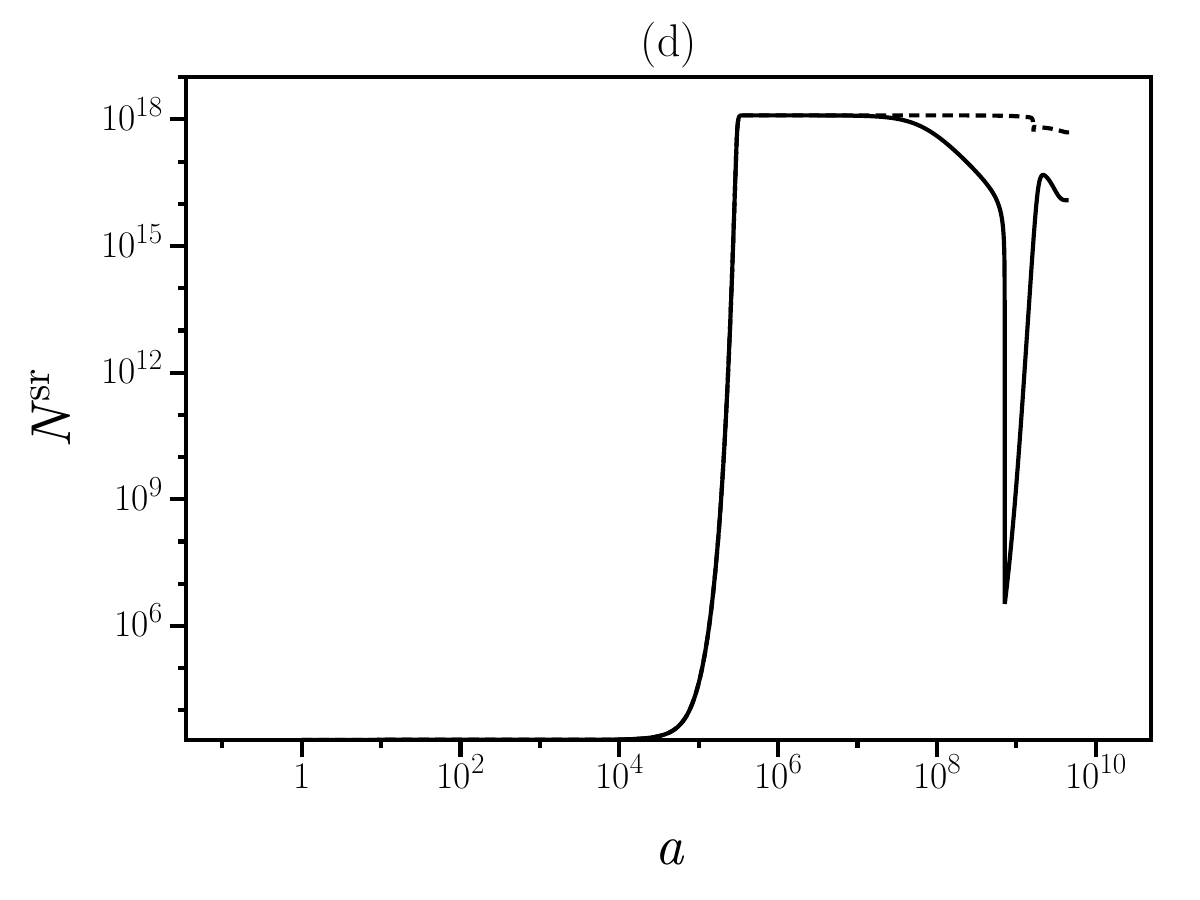}
    \includegraphics[width=0.32\linewidth]{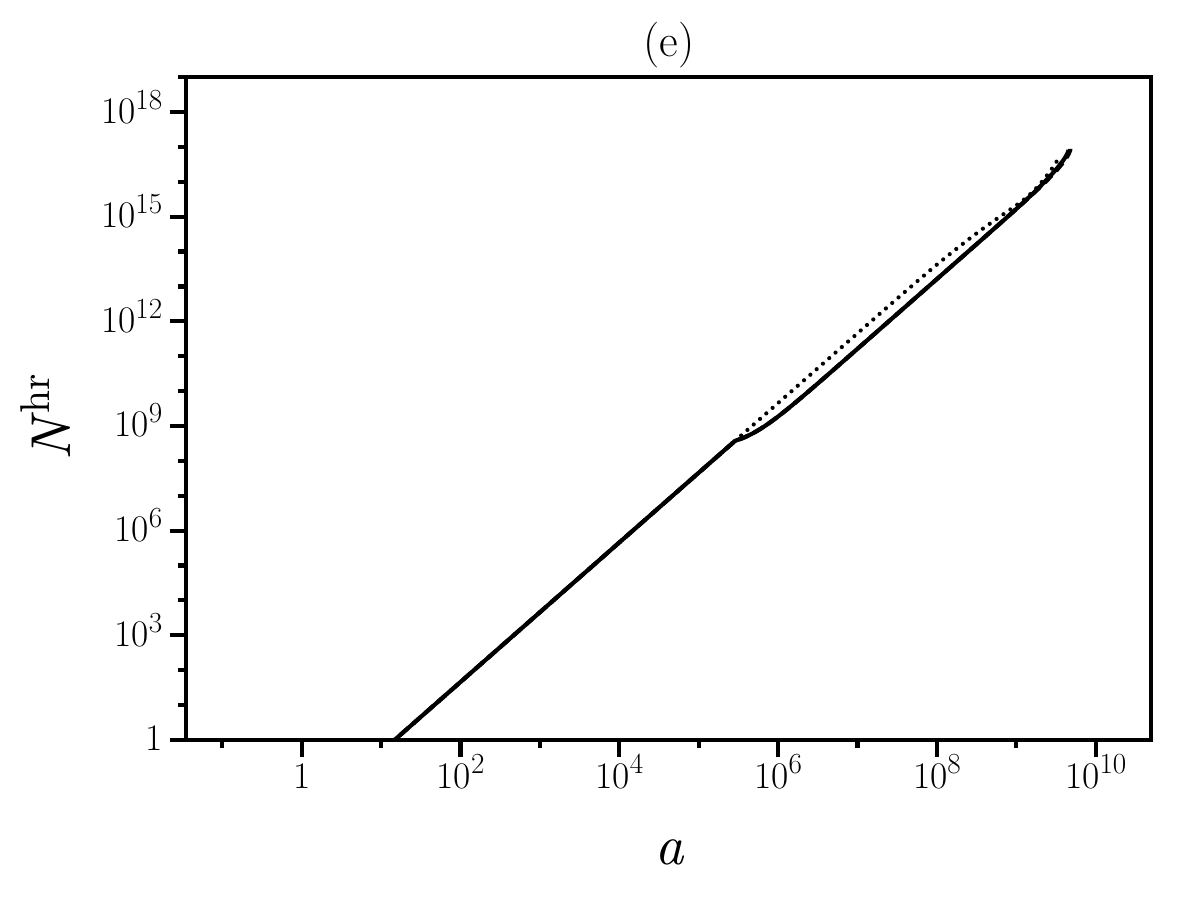}
    \caption{Evolution of key quantities in scalar superradiance, as functions of the scale factor $a$ ($a=1$ corresponds to time of PBH formation): (a) PBH mass $M$; (b) dimensionless PBH spin $a_*$; (c) number of DM particles in a comoving volume $\mathcal{N}_\mathrm{DM} \equiv n_\mathrm{DM} a^3$; (d) total occupation number $N^\mathrm{sr}$ of the superradiant cloud; and (e) number of DM particles from Hawking radiation $N^\mathrm{hr}$ . The initial parameters are PBH mass $M_\mathrm{ini} = 3 \times 10^4~\mathrm{g}$, spin $a_{*\mathrm{ini}} = 0.999$, abundance $\beta = 4.8 \times 10^{-21}$, and scalar DM particle mass $\mu = 10^9~\mathrm{GeV}$. Hawking radiation and superradiance are abbreviated as ``HR" and ``SR" respectively. The solid curves depict the scenario considering the interplay of Hawking radiation, superradiance, and GW emission, which reproduces the observed DM relic abundance ($\Omega_\mathrm{DM}h^2 \approx 0.12$). The dotted curves represent the Hawking radiation-only case, leading to the underproduction of DM ($\Omega_\mathrm{DM}h^2 \approx 0.076$). The dashed curves show the results when GW emission is ignored, leading to the overproduction of DM ($\Omega_\mathrm{DM}h^2 \approx 0.72$).}
    \label{fig:example_evo_S}
\end{figure}

\begin{figure}
    \centering
    \includegraphics[width=0.32\textwidth]{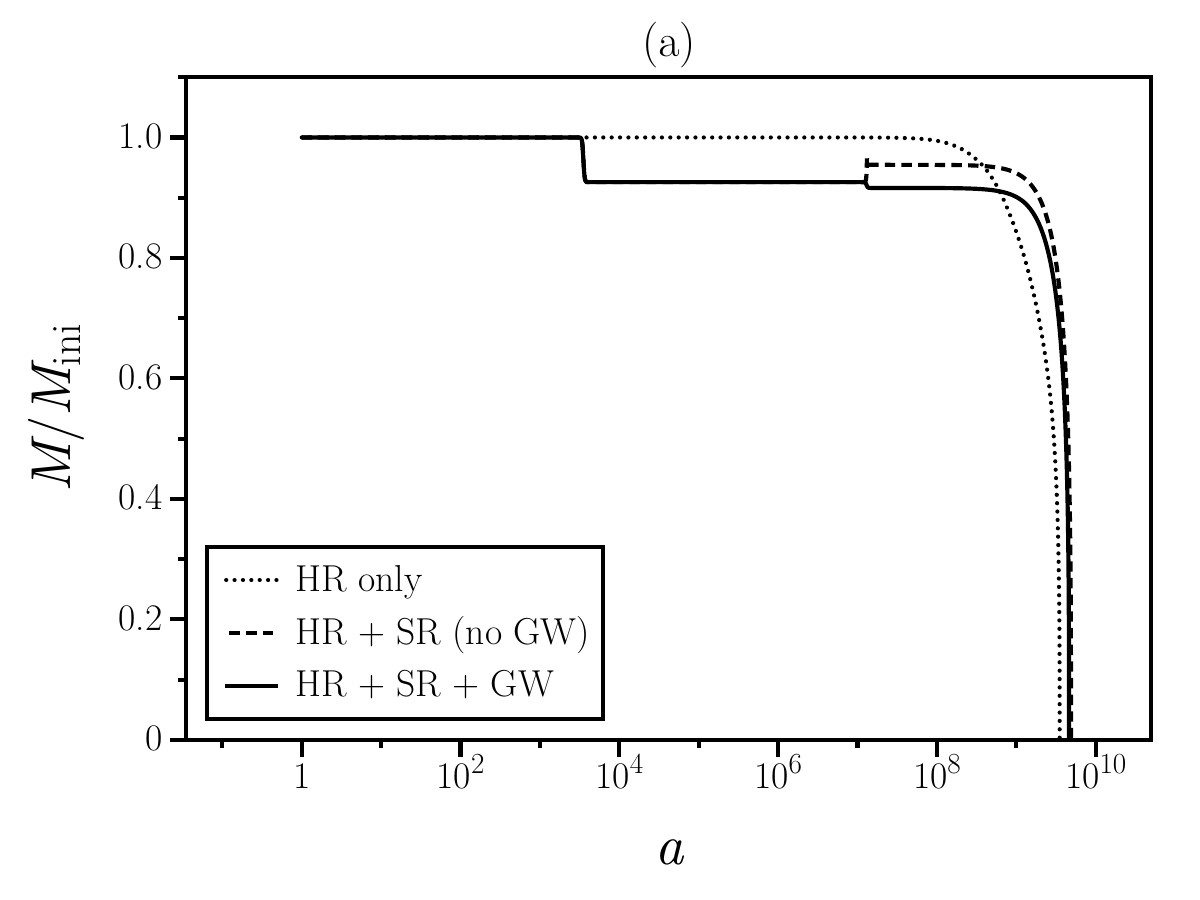}
    \includegraphics[width=0.32\textwidth]{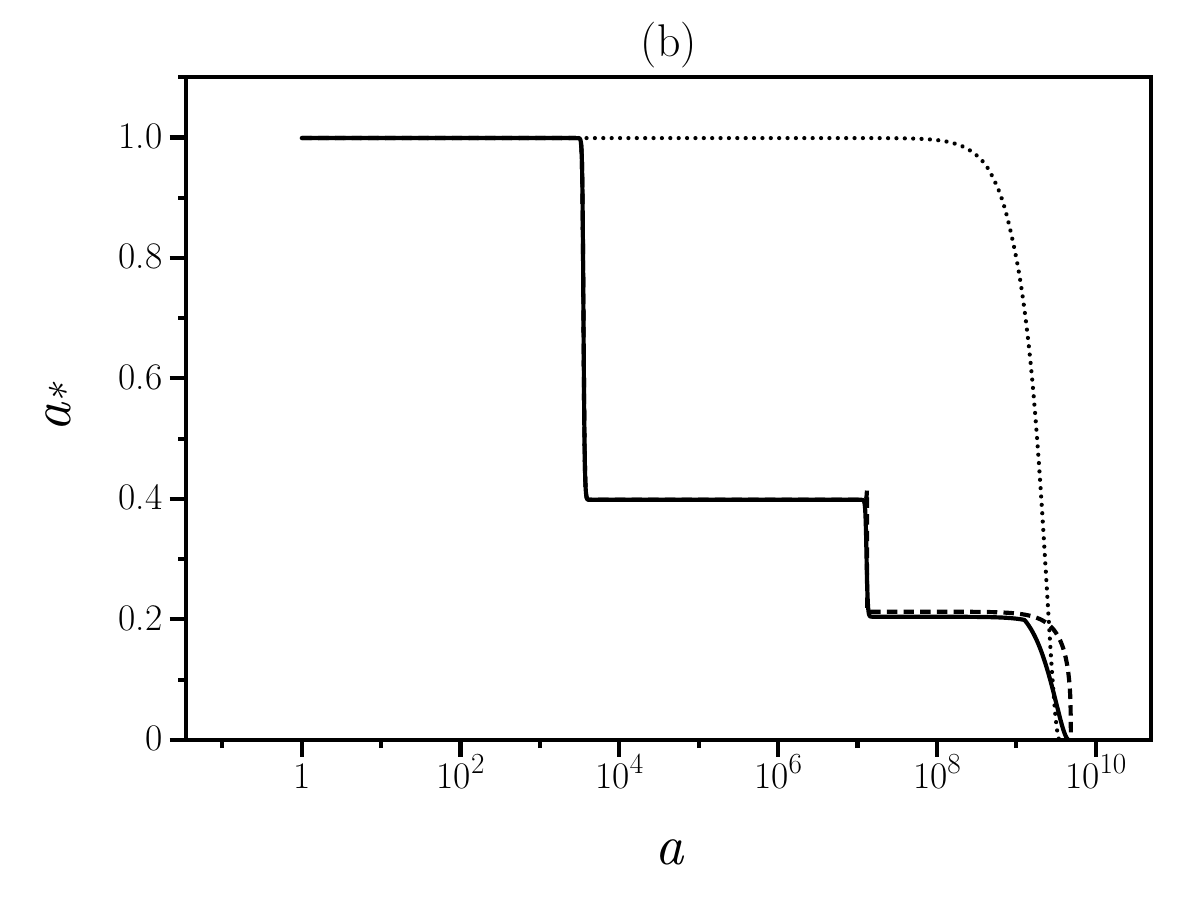}
    \includegraphics[width=0.32\textwidth]{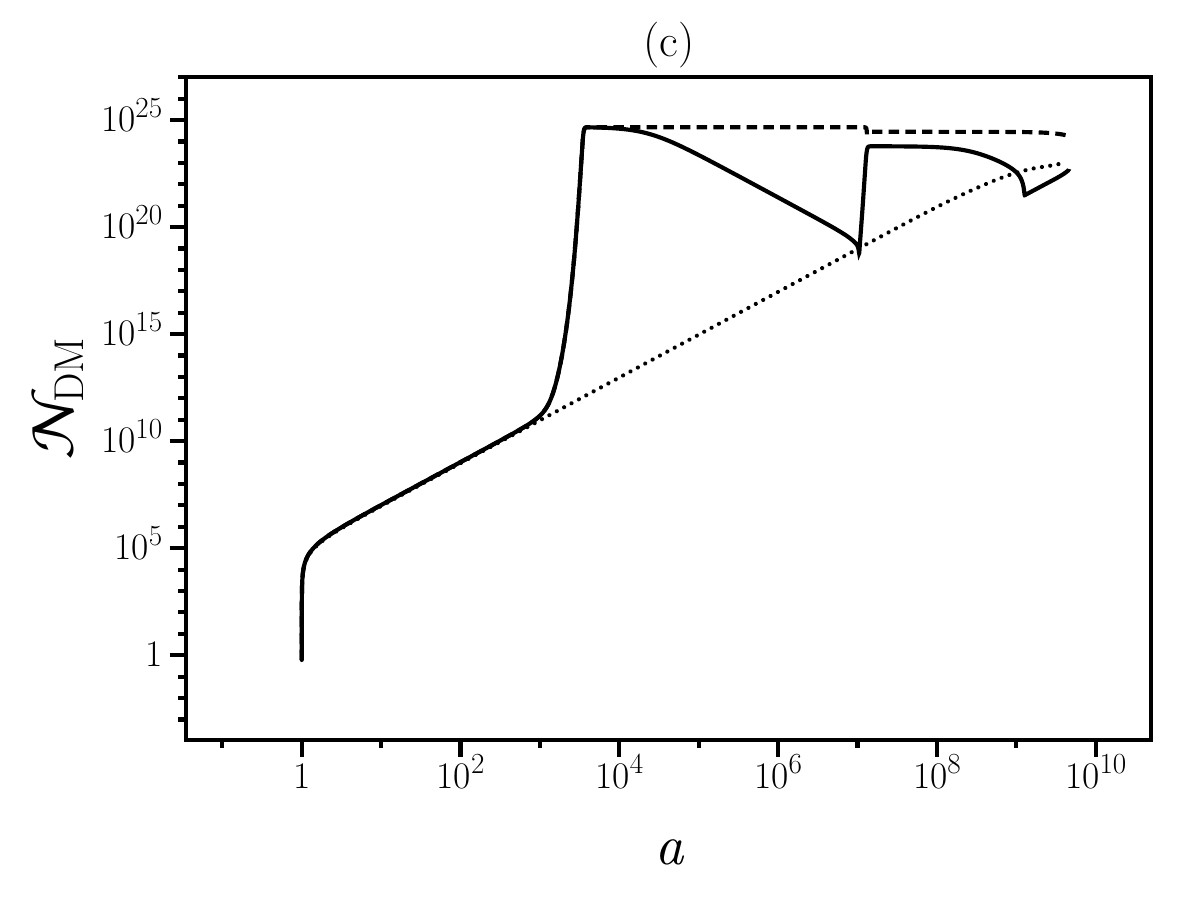}
    \includegraphics[width=0.32\linewidth]{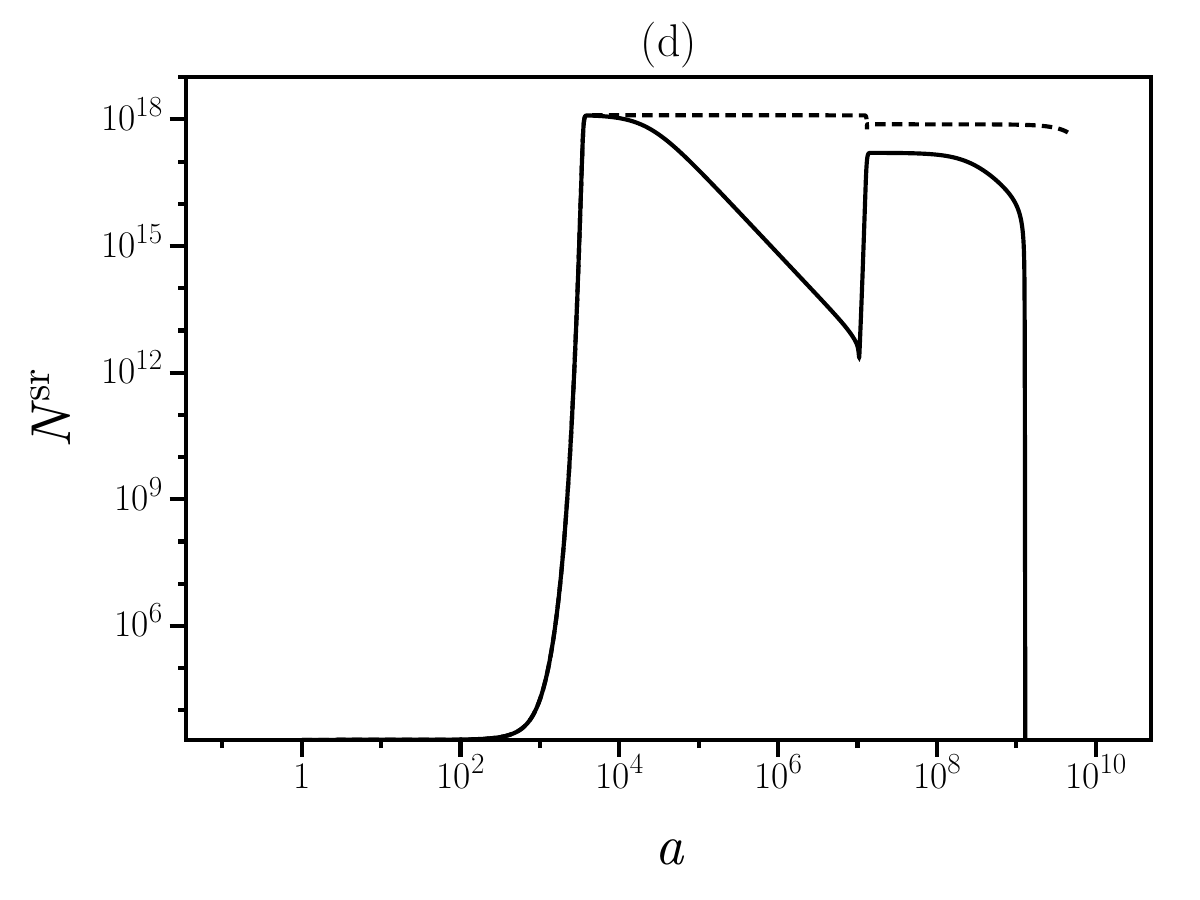}
    \includegraphics[width=0.32\linewidth]{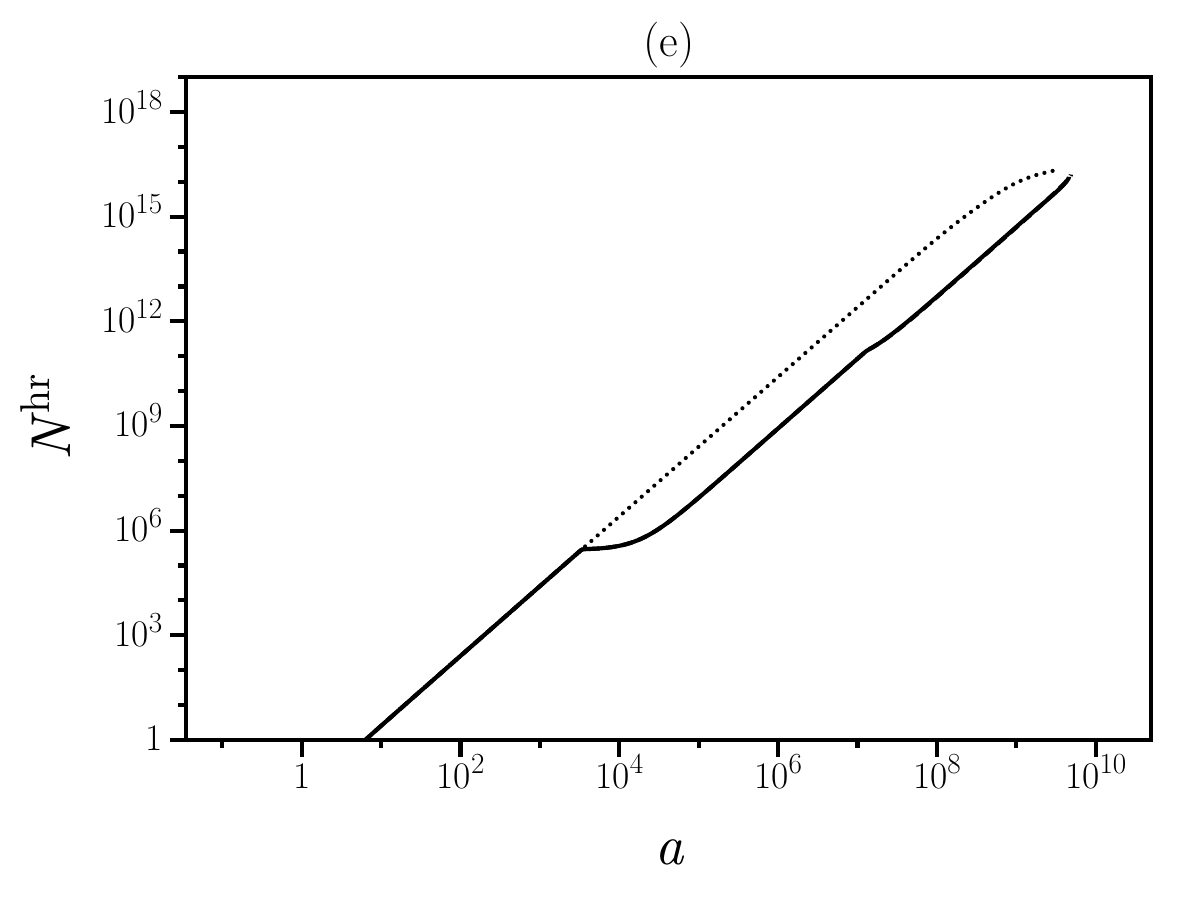}
    \caption{Evolution of key quantities in vector superradiance, as functions of the scale factor $a$ ($a=1$ corresponds to time of PBH formation): (a) PBH mass $M$; (b) dimensionless PBH spin $a_*$; (c) number of DM particles in a comoving volume $\mathcal{N}_\mathrm{DM} \equiv n_\mathrm{DM} a^3$; (d) total occupation number $N^\mathrm{sr}$ of the superradiant cloud; and (e) number of DM particles from Hawking radiation $N^\mathrm{hr}$. The initial parameters are PBH mass $M_\mathrm{ini} = 3 \times 10^4~\mathrm{g}$, spin $a_{*\mathrm{ini}} = 0.999$, abundance $\beta = 4.8 \times 10^{-21}$, and vector DM particle mass $\mu = 10^9~\mathrm{GeV}$. Hawking radiation and superradiance are abbreviated as ``HR" and ``SR" respectively. The solid curves depict the scenario considering the interplay of Hawking radiation, superradiance, and GW emission, resulting in the DM relic abundance $\Omega_\mathrm{DM}h^2 \approx 0.017$. The dotted curves represent the Hawking radiation-only case, resulting in $\Omega_\mathrm{DM}h^2 \approx 0.031$. The dashed curves show the results when GW emission is ignored, leading to $\Omega_\mathrm{DM}h^2 \approx 0.58$.}
    \label{fig:example_evo_V}
\end{figure}

In the case of superradiant scalar DM production from PBHs, the evolution of certain key quantities is illustrated in figure~\ref{fig:example_evo_S} for a benchmark set of parameters, showing (a) PBH mass $M$, (b) the dimensionless PBH spin $a_*$, (c) the number of DM particles in a comoving volume $\mathcal{N}_\mathrm{DM}\equiv n_\mathrm{DM}a^3$, (d) total occupation number $N^\mathrm{sr}\equiv\sum_iN^\mathrm{sr}_i$, and (e) Hawking emission number $N^\mathrm{hr}$. The computation uses the following parameters as initial conditions: PBH mass $M_\mathrm{ini} = 3 \times 10^4~\mathrm{g}$, dimensionless PBH spin $a_{*\mathrm{ini}} = 0.999$, abundance parameter $\beta = 4.8 \times 10^{-21}$, and spin-$0$ DM particle mass $\mu = 10^9~\mathrm{GeV}$. The initial BH mass $M_\mathrm{ini}$ determines the lifetime of the PBH, thereby setting the endpoint of the evolution and the number of superradiant DM particles left. The initial BH spin $a_{*\mathrm{ini}}$ determines the available PBH angular momentum for the superradiance process, which controls the population of superradiant DM particles. The abundance parameter $\beta$ determines the number density of the PBHs and thus directly influences the production of all particles from PBHs. The magnitude of $\mu$ roughly sets the range of $M_\mathrm{ini}$ where superradiance efficiently influences DM production, since this process becomes effective when the dimensionless mass coupling satisfies $M\mu \sim \mathcal{O}(1)$.

For the above set of parameters in the scalar DM case, when all three processes---Hawking radiation, superradiance, and GW emission---are included (solid curve), the evolution yields a DM relic abundance of $\Omega_\mathrm{DM}h^2 \approx 0.12$, consistent with the observed DM abundance. In contrast, considering only Hawking radiation (dotted curve) results in insufficient DM production ($\Omega_\mathrm{DM}h^2 \approx 0.076$), while excluding GW emission (dashed curve) leads to excessive DM production ($\Omega_\mathrm{DM}h^2 \approx 0.72$).

The maximum occupation number achieved by the first growing mode represents the highest attainable DM particle number throughout the entire evolution, but it cannot be sustained due to two primary reasons. First, GW emission continuously reduces the occupation number of the first mode. Second, the emergence and growth of the subsequent second mode can lower the BH spin below the critical threshold for the first mode, causing the first mode to become a decaying mode that is absorbed by the PBH. Consequently, the DM particles that remain after complete evaporation of the PBH mainly originate from the surviving second mode. This moderates the overall superradiant enhancement of DM production, resulting in a final $\mathcal{N}_\mathrm{DM}$ that exceeds the Hawking radiation-only scenario by merely a factor of a few, rather than by orders of magnitude.

Figure~\ref{fig:example_evo_V} presents a parallel analysis for spin-1 DM particles under identical initial conditions. The vector modes exhibit significantly shorter GW emission timescales compared to their scalar counterparts, leading to a faster annihilation of the superradiant DM particles. Additionally, the instability rates for vector modes are also substantially higher, further accelerating mode decay once they turn to their decaying phase. These two factors result in rapid depletion of the DM cloud, occurring before the complete evaporation of the PBH.

Consequently, in the vector DM case, when all three processes---Hawking radiation, superradiance, and GW emission---are considered, the evolution yields a notably lower DM relic abundance ($\Omega_\mathrm{DM}h^2 \approx 0.017$) compared to both the scenario without GW emission ($\Omega_\mathrm{DM}h^2 \approx 0.58$) and the Hawking radiation-only case ($\Omega_\mathrm{DM}h^2 \approx 0.031$). This parallel analysis emphasizes that superradiance does not universally enhance DM production. If the depletion rate of the superradiant cloud is sufficiently high, the cloud dissipates entirely before PBHs fully evaporate. In such cases, superradiance competes with Hawking radiation, causing mass that could otherwise contribute to DM production through Hawking radiation to be lost. This mechanism explains why, for vector DM, the resulting DM abundance is even lower than in the Hawking radiation-only scenario, illustrating that superradiance can counter-intuitively suppress rather than enhance DM production.

\begin{figure}[t]
\centering
\includegraphics[width=0.49\textwidth]{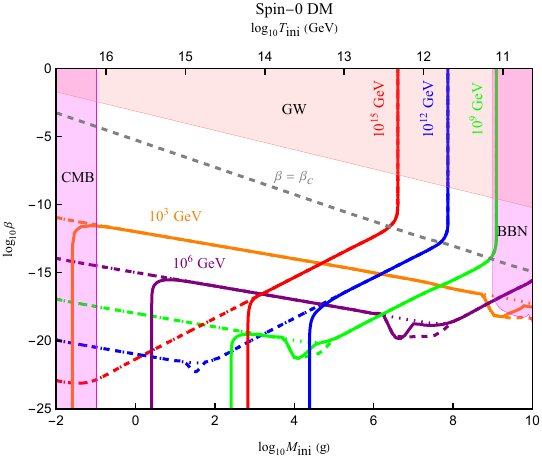}
\includegraphics[width=0.49\textwidth]{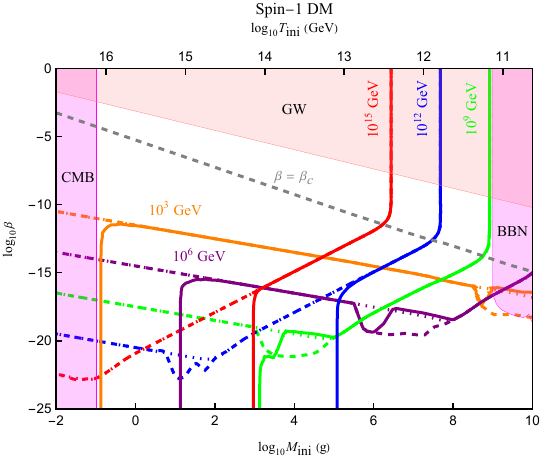}
\caption{Viable parameter space for DM production in the $M_\mathrm{ini}$-$\beta$ plane. Contours show parameter combinations yielding the observed DM relic density ($\Omega_\mathrm{DM}h^2=0.12$) for a nearly extremal initial PBH spin ($a_{*\mathrm{ini}}=0.999$). Shaded regions indicate excluded parameter space from BBN (eq.\eqref{eq:BBN}), CMB (eq.\eqref{eq:CMB}), and GW bounds (eq.\eqref{eq:GW}). The grey dashed line marks the threshold for a PBH-dominated era to present (eq.\eqref{eq:PBHD}). For DM masses from $10^3$ to $10^{15}$ GeV, we compare three scenarios: PBH evolution with superradiance including GW emission and also the effect of gravitational UV freeze-in (solid curves), superradiance without GW emission (dashed curves), and pure Hawking radiation (dotted curves).}
\label{fig:scan}
\end{figure}

Figure~\ref{fig:scan} illustrates the viable parameter space for DM production in the $M_\mathrm{ini}$-$\beta$ plane, where contours represent parameter combinations that reproduce the observed DM relic density, assuming a nearly extremal initial PBH spin $a_{*\mathrm{ini}}=0.999$. The parameter space is constrained by three physical bounds shown as shaded regions: the BBN constraints from eq.~\eqref{eq:BBN}, the CMB limits from eq.~\eqref{eq:CMB}, and the GW bounds from eq.~\eqref{eq:GW}. A grey dashed line, derived from eq.~\eqref{eq:PBHD}, shows the threshold for an early PBH-dominated era. In the region above this line, where the PBH domination occurs, the final DM abundance depends solely on the initial PBH mass. We examine DM production across masses ranging from $10^3$ to $10^{15}$~GeV (depicted in different colors), comparing three distinct combinations of mechanisms. The solid curves represent PBH evolutions with superradiance and GW emission, and also include the effects of gravitational UV freeze-in (see section~\ref{sec:UV} for details). The dashed curves represent PBH evolutions with superradiance but without considering GW emission, and the dotted curves illustrate the case of pure Hawking radiation.

The scalar DM case (left panel) demonstrates consistent enhancement of DM production through superradiant instabilities relative to pure Hawking radiation, though this enhancement is significantly moderated by GW emission from the superradiant cloud. In contrast, vector DM (right panel) exhibits more complex phenomenology across the parameter space: the relative positions of the solid and dotted lines reveal distinct scenarios where superradiance can either enhance or suppress DM production compared to the Hawking radiation-only case. These results demonstrate how differences in the superradiant evolution influence DM production outcomes for scalar and vector modes across the parameter space.

\section{Gravitational UV freeze-in}\label{sec:UV}

The gravitational UV freeze-in mechanism describes the production of DM particles through graviton-mediated $2 \rightarrow 2$ annihilation of SM particles. This process provides an irreducible contribution to DM genesis, independent of other production channels. The evolution of the DM number density $n^\mathrm{uv}_\mathrm{DM}$ generated through UV freeze-in is governed by the Boltzmann equation~\cite{Bernal:2020ili,Tang:2017hvq,Garny:2015sjg,Giudice:2000ex}:
\begin{equation}
    \frac{\mathrm{d}n^\mathrm{uv}_\mathrm{DM}}{\mathrm{d}t} + 3Hn^\mathrm{uv}_\mathrm{DM} =
    \left(n_\mathrm{DM}^{\mathrm{eq}}\right)^2\langle\sigma v\rangle,
\end{equation}
where $n_\mathrm{DM}^{\mathrm{eq}}$ represents the equilibrium number density of DM particles at temperature $T$, and $\langle\sigma v\rangle$ denotes the thermally-averaged cross-section for the annihilation process. The source term on the right-hand side can be parameterized as:
\begin{equation}
    \left(n_\mathrm{DM}^{\mathrm{eq}}\right)^2\langle\sigma v\rangle =
    \delta\frac{\left(8\pi\right)^2 T^8}{M_\mathrm{Pl}^4}.
\end{equation}
The coupling coefficient $\delta$ depends on the DM spin state: $\delta = 1.9 \times 10^{-4}$ for scalar DM, and $\delta = 2.3 \times 10^{-3}$ for vector DM~\cite{Bernal:2018qlk}.

For DM masses exceeding the reheating temperature $T_\mathrm{rh}$ while remaining below the maximum temperature of the SM thermal bath, the UV freeze-in yield experiences Boltzmann suppression by a factor $\exp(-2\mu/T_\mathrm{rh})$~\cite{Giudice:2000ex}. In this work, we consider PBH formation during the early radiation-dominated era after reheating, implying $T_\mathrm{rh} \geq T_\mathrm{ini}$. Under these conditions, the DM yield from UV freeze-in reaches its minimum value when $T_\mathrm{rh} = T_\mathrm{ini}$. Thus we can make a conservative bound on the initial temperature required for UV freeze-in to generate the observed DM abundance.

The solid curves in figure~\ref{fig:scan} illustrate the impact of UV freeze-in. Note that the upper axes of figure~\ref{fig:scan} show the corresponding initial cosmological temperature $T_\mathrm{ini}$, and the relation $T_\mathrm{ini} \propto M_\mathrm{ini}^{-1/2}$ can be derived from eq.~\eqref{eq:Mini}. The solid lines deviate from the dashed lines at small $M_\mathrm{ini}$ values. This behavior can be traced to our requirement $T_\mathrm{rh} \geq T_\mathrm{ini}$. Specifically, smaller PBHs form at higher initial temperatures, which in turn requires a higher reheating temperature $T_\mathrm{rh}$ since we assume PBHs are formed after reheating. Higher $T_\mathrm{rh}$ enhances the DM yield from UV freeze-in, and eventually this contribution becomes dominant for small $M_\mathrm{ini}$. The conservative scenario of minimum DM yield from UV freeze-in with $T_\mathrm{rh} = T_\mathrm{ini}$ is assumed in figure~\ref{fig:scan}. Larger contribution from UV freeze-in will diminish more parameter region
available for the superradiant production considered in this work.

It is also worth noting that the inclusion of superradiance has a negligible influence on UV freeze-in effects for two key reasons. First, in regions where UV freeze-in efficiently produces the observed DM relic abundance, the values of $\beta$ are sufficiently low, so that the DM production from both Hawking radiation and superradiance becomes negligible. Second, as can be observed from figures~\ref{fig:example_evo_S} and \ref{fig:example_evo_V}, and eq.~\eqref{eq:SMevo}, superradiance minimally affects both the PBH lifetime and the universe's temperature evolution.

\section{Discussion and conclusions}\label{sec:conclusion}

In this work, we have investigated heavy bosonic DM production from PBHs in the early universe, incorporating multiple gravitational mechanisms: Hawking radiation, superradiant instabilities, and gravitational UV freeze-in. We demonstrated that incorporating multiple-mode effects and GW emission
leads to revised estimates of the DM yield. In scenarios involving two sequential superradiant modes, the dominant contribution to the superradiant DM yield typically arises from the second mode. This occurs because the first mode generally fails to persist to the end of PBH evaporation due to energy loss via GW emission and mode decay.

For scalar DM, while superradiant instabilities can still enhance production compared to pure Hawking radiation, the resulting enhancement factor is moderated. In the case of vector DM, for a certain range of parameters, superradiance can act competitively in DM production, which results from the higher instability rates and greater GW emission power of vector modes relative to scalar modes. Consequently, the rapid formation and dissipation of vector superradiant clouds lead to redirection of PBH mass, suppressing the final DM relic abundance below those achieved through Hawking radiation alone.

Although our analysis assumes a near-extreme initial dimensionless PBH spin of $a_{*\mathrm{ini}} = 0.999$, adopting a lower spin value does not qualitatively alter the evolutionary scenario described here. It simply reduces the viable parameter space capable of generating the observed DM relic abundance.

We have included gravitational UV freeze-in in our analysis. In regions of parameter space where gravitational UV freeze-in efficiently generates the observed DM abundance, the contributions from both Hawking radiation and superradiance become subdominant. Additionally, since superradiance produces minimal modifications to the universe's temperature evolution, it does not significantly alter the process of gravitational UV freeze-in.

Potential observable consequences of superradiant DM production from PBHs are limited to the DM relic abundance and the GW emissions from the superradiant cloud. However, the associated GW frequencies lie in a very high-frequency domain~\cite{Aggarwal:2020olq}. Consequently, direct observational confirmation of PBH-mediated DM genesis is challenging. However, incorporating superradiance effects modifies the constraints on parameter space associated with DM production, and our work has refined this modified parameter space further. This can provide an indirect method for evaluating the viability and cosmological significance of PBH-mediated DM genesis within the broader context of DM production mechanisms.

Our work can be extended in a number of directions. If PBHs exist, it is likely that the PBH mass function exhibits a
more or less extended spectrum. It is therefore motivated to
investigate the impact of an extended PBH mass function on
superradiant DM production. Moreover, although we considered DM with pure gravitational interaction, it is
evident that the issues considered in this work might persist
in more complicated DM models featuring DM interacting with
other SM particles and/or DM self-interactions. It is thus
important to carry out the analysis for these more complicated
scenarios. Last but not least, it is interesting to consider
the impact of alternative cosmic expansion histories on
superradiant DM production. We believe that the analysis performed here and in the extended studies might have important
consequences for a number of scenarios proposed in the literature, such as those in refs.~\cite{Ghoshal:2023fno,Manno:2025dhw,Jiang:2025blz}.

\begin{acknowledgments}
The authors would like to thank Yin-da Guo for helpful discussions. This work is supported by the National Natural Science Foundation of China (Grants Nos.~12473001, 11975072, 11875102, and 11835009), the National SKA Program of China ( Grants Nos.~2022SKA0110200 and 2022SKA0110203), the National 111 Project (Grant No. B16009), and the China Manned Space Program with grant no. CMS-CSST-2025-A02. C. Zhang is supported by the Joint Fund of Natural Science Foundation of Liaoning Province (Grant No.~2023-MSBA-067) and the Fundamental Research Funds for the Central Universities (Grant No.~N2405011). S-S. Bao and H. Zhang are supported by the National Natural Science Foundation of China (Grant No.~12447105).
\end{acknowledgments}

\appendix
\section{Superradiant modes}\label{ap:sr_modes}

This appendix explains on our rationale for selecting the specific superradiant modes $\left|211\right\rangle$ and $\left|322\right\rangle$ for scalar fields, and $\left|1011\right\rangle$ and $\left|2122\right\rangle$ for vector fields in our analysis. We also explain
the implementation of a cutoff in the gravitational coupling
in the superradiance computation.

\subsection{Instability rate formalism}

The theoretical framework for superradiant instabilities of massive bosonic fields (spin-$0$, spin-$1$, and spin-$2$) around Kerr black holes has been extensively developed in the literature~\cite{Brito:2015oca}. When a bosonic field propagates in the vicinity of a Kerr black hole with horizon angular velocity $\Omega_\mathrm{H}$, it can form quasibound states characterized by angular frequencies $\omega$ and magnetic quantum numbers $m$. These states exhibit exponential growth when they satisfy the superradiance condition $\omega_R < m \Omega_\mathrm{H}$, where $\omega_R$ represents the real part of the complex eigenfrequency $\omega = \omega_R + i \omega_I$. For the scalar case, $\omega_R$ is given by~\cite{Baumann:2019eav}:
\begin{equation}
    \omega_R=\mu\left(1-\frac{\left(M\mu\right)^2}{2 n^2}-\frac{\left(M\mu\right)^4}{8 n^4}+\frac{f_{n l}}{n^3} \left(M\mu\right)^4+\frac{h_{l}}{n^3} {a}_* m \left(M\mu\right)^5+\cdots\right),
\end{equation}
with
\begin{equation}
    f_{n l} \equiv-\frac{6}{2 l+1}+\frac{2}{n} \text{ and }
    h_{l} \equiv \frac{16}{2 l(2 l+1)(2 l+2)} .
\end{equation}
For the vector case, $\omega_R$ reads
\begin{equation}
    \omega_R=\mu\left(1-\frac{\left(M\mu\right)^2}{2 n^2}-\frac{\left(M\mu\right)^4}{8 n^4}+\frac{f_{n l j}}{n^3} \left(M\mu\right)^4+\frac{h_{l j}}{n^3} {a}_* m \left(M\mu\right)^5+\cdots\right),
\end{equation}
with
\begin{equation}
    \begin{aligned}
    f_{n l j} & =-\frac{4(6 l j+3 l+3 j+2)}{(l+j)(l+j+1)(l+j+2)}+\frac{2}{n}, \\
    h_{l j} & =\frac{16}{(l+j)(l+j+1)(l+j+2)}.
    \end{aligned}
\end{equation}
The imaginary part $\omega_I$ determines the growth rate of the field amplitude, while the superradiant growth rate of the occupation number is given by $\Gamma \equiv 2 \omega_I$. This factor of 2 arises naturally from the quadratic relationship between occupation number and field amplitude. When the superradiance condition is no longer satisfied, $\Gamma$ becomes non-positive and growth ceases.

In the analytically tractable regime where $M\mu \ll 1$, the instability rate for a scalar mode $\left|nlm\right\rangle$ can be expressed as~\cite{Brito:2015oca,Baumann:2019eav,Bao:2022hew}:
\begin{equation}\label{eq:fullsrS}
\Gamma_{nlm}^{\mathrm{sr}} = 2\times \frac{2r_+}{M} C_{nl} \prod_{k=1}^{l}\left(k^2\left(1-a_*^2\right)+\left(a_* m-2 r_{+} \omega_R\right)^2\right) \left(m \Omega_{\mathrm{H}}-\omega_R\right)\ \left(M\mu\right)^{4 l+5}
\end{equation}
where $C_{nl}$ represents the coefficient defined in the main text. Under the hydrogenic approximation for bound states, $\omega_R \approx \mu$, which allows us to derive the simplified expression presented in eq.~\eqref{eq:GammasrS} for small values of $l$. For vector modes $\left|nljm\right\rangle$, the corresponding instability rate reads:
\begin{equation}\label{eq:fullsrV}
\Gamma_{nljm}^{\mathrm{sr}} = 2\times \frac{2r_+}{M} C_{nlj} \prod_{k=1}^{j}\left(k^2\left(1-a_*^2\right)+\left(a_* m-2 r_{+} \omega_R\right)^2\right) \left(m \Omega_{\mathrm{H}}-\omega_R\right)\ \left(M\mu\right)^{2l+2j+5}
\end{equation}
Similarly, this expression reduces to the approximate form given in eq.~\eqref{eq:GammasrV}.

\subsection{Mode selection criteria}

Our selection of specific superradiant modes for analysis follows the principles based on growth efficiency and evolutionary dynamics:

First, we prioritize modes with the highest growth rates. From equations \eqref{eq:fullsrS} and \eqref{eq:fullsrV}, it is evident that modes with the lowest total angular momentum number ($l$ for scalar fields or $j$ for vector fields) grow most rapidly. Furthermore, among modes sharing the same (but not too large) $l$ or $j$, those with the lowest principal quantum number $n$ exhibit superior growth rates. Therefore, we identify $\left|211\right\rangle$ for scalar fields and $\left|1011\right\rangle$ for vector fields as our primary modes, setting $m=1$ to ensure the superradiance condition be satisfied.

For the secondary modes, we select those with the next lowest $l$ or $j$: $\left|322\right\rangle$ for scalar fields and $\left|2122\right\rangle$ for vector fields. As demonstrated by equations \eqref{eq:fullsrS} and \eqref{eq:fullsrV}, these secondary modes grow approximately an order of magnitude slower than their primary counterparts at comparable $M\mu$ values. This growth rate difference creates a natural sequential evolution pattern, where the secondary mode becomes significant only after substantial evolution of the primary mode. Given the range of PBH initial masses considered in our study, this two-mode approach effectively captures the essential dynamics throughout the entire PBH lifetime, so the inclusion of additional modes is unnecessary for our analysis.

\subsection{Validity and numerical implementation}

The analytic expressions in equations \eqref{eq:fullsrS} and \eqref{eq:fullsrV} demonstrate excellent agreement with numerical results reported in the literature~\cite{Dolan:2007mj,Baumann:2019eav,Dolan:2018dqv,Baryakhtar:2017ngi}. We employ these equations to model the superradiant evolution in our system. To ensure both the validity of analytic approximations and the numerical stability, we implement a dimensionless mass coupling cutoff at $M\mu = \alpha_c$, where $\alpha_c < 1$. This means superradiance effects are only considered when $M\mu \leq \alpha_c$.

Since Hawking radiation continuously reduces the PBH mass during evolution, the value of $\alpha_c$ influences the timing of superradiance onset. A more conservative (smaller) value of $\alpha_c$ would delay the initiation of superradiance and consequently alter the system's evolutionary trajectory. However, the DM production from superradiance remains quantitatively constant across reasonable variations of $\alpha_c$, provided that the cutoff encompasses the primary range of $M\mu$ where superradiance operates most efficiently. This robustness validates the reliability of our main results and conclusions.

\section{Other sources of GW emission}\label{ap:trans_GWs}

In this work, we have investigated the GW emission arising from annihilations between pairs of superradiant particles within the same mode. This channel represents the sole mechanism when considering the evolution of a single superradiant mode exclusively. However, our investigation extends to the case involving the coexistence of two superradiant modes, where there exist additional GW emission channels. In this appendix
we give justification for neglecting them in our study.

When multiple modes are taken into account, there exist energy-level transitions between different modes, which yield monochromatic GWs with frequency $\omega_\mathrm{GW} = \Delta \omega$, where
\begin{equation}
    \Delta \omega = \frac{\mu\left(M\mu\right)^{2}}{2}\left(\frac{1}{n_g^2}-\frac{1}{n_e^2}\right)
\end{equation}
represents the energy difference between the ``ground" state $g$ and the ``excited" state $e$. In systems where two modes dominate the superradiance process—one in the ``ground" state and another in the ``excited" state—the evolution of their respective occupation numbers $N_{e, g}$ is governed by their individual instability rates and further modified by transitions from excited to ground states that emit GWs. The rate of change in occupation number due to these transitions can be formulated as:
\begin{equation}
\frac{d N_\mathrm{tr}}{d t} = \frac{P^\mathrm{GW}_\mathrm{tr}}{\Delta \omega},
\end{equation}
where $P^\mathrm{GW}_\mathrm{tr}$ denotes the GW power emitted during transitions. Through the introduction of the transition rate parameter $\Gamma_\mathrm{tr}$, this relationship can alternatively be expressed as:
\begin{equation}
    \frac{d N_\mathrm{tr}}{d t}=\Gamma_\mathrm{tr} N_e N_g.
\end{equation}
Since the GW wavelength from transitions typically exceeds the characteristic scale of the superradiant cloud, the quadrupole formalism~\cite{Weinberg:1972kfs} provides an appropriate framework for calculating this emission. Therefore, we can write~\cite{Arvanitaki:2010sy,Arvanitaki:2014wva}:
\begin{equation}
    \left.\frac{d N_\mathrm{tr}}{d t}\right|_{\mathrm{quadr}}=N_e N_g \frac{2 \Delta \omega^5}{5} I_{ij} I^{ij},
\end{equation}
where $I_{ij}$ represents the transition mass quadrupole moment, which can be approximated as
\begin{equation}
    I_{ij} I^{ij} \sim \mu^2 r_c^4,
\end{equation}
with the superradiant cloud radius $r_c \sim M {{n}^2}/{(M\mu)^2}$.

GWs from energy-level transitions become relevant mainly for two modes with the same $l$ and $m$ but differing in $n$, especially for sufficiently high $l$ cases. For instance, the GW emission rate from the transition $\left|644\right\rangle \rightarrow \left|544\right\rangle$ is estimated as~\cite{Arvanitaki:2010sy,Arvanitaki:2014wva}
\begin{equation}
    P^\mathrm{GW}_\mathrm{tr}\left(\left|644\right\rangle \rightarrow \left|544\right\rangle\right) \sim \mathcal{O}\left(10^{-7}\right) \frac{M^\mathrm{sr}_{644} M^\mathrm{sr}_{544}}{M^2} \left(M\mu\right)^8.
\end{equation}
On the other hand, the transition $\left|322\right\rangle \rightarrow \left|211\right\rangle$ yields a considerably smaller rate due to suppression by higher-order terms in $M\mu$~\cite{Arvanitaki:2010sy}:
\begin{equation}
    \begin{aligned}
        P^\mathrm{GW}_\mathrm{tr}(\left|322\right\rangle \rightarrow \left|211\right\rangle) &\approx N^\mathrm{sr}_{322} N^\mathrm{sr}_{211} \frac{5717 \times 2^8 \left(M\mu\right)^{14}}{3^5 5^{11} 7^3 M^4} \\
        &\approx 3.6 \times 10^{-7} \frac{M^\mathrm{sr}_{322}  M^\mathrm{sr}_{211}}{M^2} \left(M\mu\right)^{12}.
    \end{aligned}
\end{equation}
The suppression arises because quadrupole transition between $\left|211\right\rangle$ and $\left|322\right\rangle$ are forbidden by selection rules. Additionally, as the occupation numbers $N^\mathrm{sr}_{211}$ and $N^\mathrm{sr}_{322}$ reach their maximum values sequentially rather than simultaneously, this GW emission channel becomes even less relevant. In other words, since the two coexisting modes we consider can only arise one after another, and cannot coexist in their respective maximum occupation number. This consideration also applies to
the annihilation of two different modes into gravitions.

In summary, the strengths of the two cross-mode effects are proportional to the product of the two respective occupation numbers, and thus are mostly not comparable to the annihilations of pairs in the same mode. Therefore, we neglect these effects in this work and reasonably focus only on GW emissions from annihilations of particle pairs in the same mode.

\bibliography{sdm_vb}

\providecommand{\href}[2]{#2}\begingroup\raggedright\begin{thebibliography}{100}

\bibitem{vanAlbada:1984js}
T.~S. van Albada, J.~N. Bahcall, K.~Begeman and R.~Sancisi, \emph{{The
  Distribution of Dark Matter in the Spiral Galaxy {NGC}-3198}},
  \href{https://doi.org/10.1086/163375}{\emph{Astrophys. J.} {\bfseries 295}
  (1985) 305--313}.

\bibitem{Treu:2004wt}
T.~Treu and L.~V.~E. Koopmans, \emph{{Massive dark - matter halos and evolution
  of early - type galaxies to z = 1}},
  \href{https://doi.org/10.1086/422245}{\emph{Astrophys. J.} {\bfseries 611}
  (2004) 739--760}, [\href{https://arxiv.org/abs/astro-ph/0401373}{{\ttfamily
  astro-ph/0401373}}].

\bibitem{DES:2021wwk}
{\scshape DES} collaboration, T.~M.~C. Abbott et~al., \emph{{Dark Energy Survey
  Year 3 results: Cosmological constraints from galaxy clustering and weak
  lensing}}, \href{https://doi.org/10.1103/PhysRevD.105.023520}{\emph{Phys.
  Rev. D} {\bfseries 105} (2022) 023520},
  [\href{https://arxiv.org/abs/2105.13549}{{\ttfamily 2105.13549}}].

\bibitem{Planck:2018vyg}
{\scshape Planck} collaboration, N.~Aghanim et~al., \emph{{Planck 2018 results.
  VI. Cosmological parameters}},
  \href{https://doi.org/10.1051/0004-6361/201833910}{\emph{Astron. Astrophys.}
  {\bfseries 641} (2020) A6},
  [\href{https://arxiv.org/abs/1807.06209}{{\ttfamily 1807.06209}}].

\bibitem{Cirelli:2024ssz}
M.~Cirelli, A.~Strumia and J.~Zupan, \emph{{Dark Matter}},
  \href{https://arxiv.org/abs/2406.01705}{{\ttfamily 2406.01705}}.

\bibitem{Profumo:2017hqp}
S.~Profumo, \emph{{An Introduction to Particle Dark Matter}}.
\newblock World Scientific, 2017,
  \href{https://doi.org/10.1142/q0001}{10.1142/q0001}.

\bibitem{Mambrini:2021cwd}
Y.~Mambrini, \emph{{Particles in the Dark Universe. A Student\textquoteright{}s
  Guide to Particle Physics and Cosmology}}.
\newblock Springer, 2021,
  \href{https://doi.org/10.1007/978-3-030-78139-2}{10.1007/978-3-030-78139-2}.

\bibitem{Arcadi:2017kky}
G.~Arcadi, M.~Dutra, P.~Ghosh, M.~Lindner, Y.~Mambrini, M.~Pierre et~al.,
  \emph{{The waning of the WIMP? A review of models, searches, and
  constraints}},
  \href{https://doi.org/10.1140/epjc/s10052-018-5662-y}{\emph{Eur. Phys. J. C}
  {\bfseries 78} (2018) 203},
  [\href{https://arxiv.org/abs/1703.07364}{{\ttfamily 1703.07364}}].

\bibitem{Roszkowski:2017nbc}
L.~Roszkowski, E.~M. Sessolo and S.~Trojanowski, \emph{{WIMP dark matter
  candidates and searches\textemdash{}current status and future prospects}},
  \href{https://doi.org/10.1088/1361-6633/aab913}{\emph{Rept. Prog. Phys.}
  {\bfseries 81} (2018) 066201},
  [\href{https://arxiv.org/abs/1707.06277}{{\ttfamily 1707.06277}}].

\bibitem{Marsh:2015xka}
D.~J.~E. Marsh, \emph{{Axion Cosmology}},
  \href{https://doi.org/10.1016/j.physrep.2016.06.005}{\emph{Phys. Rept.}
  {\bfseries 643} (2016) 1--79},
  [\href{https://arxiv.org/abs/1510.07633}{{\ttfamily 1510.07633}}].

\bibitem{Irastorza:2018dyq}
I.~G. Irastorza and J.~Redondo, \emph{{New experimental approaches in the
  search for axion-like particles}},
  \href{https://doi.org/10.1016/j.ppnp.2018.05.003}{\emph{Prog. Part. Nucl.
  Phys.} {\bfseries 102} (2018) 89--159},
  [\href{https://arxiv.org/abs/1801.08127}{{\ttfamily 1801.08127}}].

\bibitem{Elahi:2014fsa}
F.~Elahi, C.~Kolda and J.~Unwin, \emph{{UltraViolet Freeze-in}},
  \href{https://doi.org/10.1007/JHEP03(2015)048}{\emph{JHEP} {\bfseries 03}
  (2015) 048}, [\href{https://arxiv.org/abs/1410.6157}{{\ttfamily 1410.6157}}].

\bibitem{Albouy:2022cin}
G.~Albouy et~al., \emph{{Theory, phenomenology, and experimental avenues for
  dark showers: a Snowmass 2021 report}},
  \href{https://doi.org/10.1140/epjc/s10052-022-11048-8}{\emph{Eur. Phys. J. C}
  {\bfseries 82} (2022) 1132},
  [\href{https://arxiv.org/abs/2203.09503}{{\ttfamily 2203.09503}}].

\bibitem{Ali-Haimoud:2017rtz}
Y.~Ali-Ha\"\i{}moud, E.~D. Kovetz and M.~Kamionkowski, \emph{{Merger rate of
  primordial black-hole binaries}},
  \href{https://doi.org/10.1103/PhysRevD.96.123523}{\emph{Phys. Rev. D}
  {\bfseries 96} (2017) 123523},
  [\href{https://arxiv.org/abs/1709.06576}{{\ttfamily 1709.06576}}].

\bibitem{Bird:2022wvk}
S.~Bird et~al., \emph{{Snowmass2021 Cosmic Frontier White Paper: Primordial
  black hole dark matter}},
  \href{https://doi.org/10.1016/j.dark.2023.101231}{\emph{Phys. Dark Univ.}
  {\bfseries 41} (2023) 101231},
  [\href{https://arxiv.org/abs/2203.08967}{{\ttfamily 2203.08967}}].

\bibitem{Zhang:2023tfv}
C.~Zhang and X.~Zhang, \emph{{Gravitational capture of magnetic monopoles by
  primordial black holes in the early universe}},
  \href{https://doi.org/10.1007/JHEP10(2023)037}{\emph{JHEP} {\bfseries 10}
  (2023) 037}, [\href{https://arxiv.org/abs/2302.07002}{{\ttfamily
  2302.07002}}].

\bibitem{Zhu:2023gmx}
Q.-H. Zhu, Z.-C. Zhao, S.~Wang and X.~Zhang, \emph{{Unraveling the early
  universe\textquoteright{}s equation of state and primordial black hole
  production with PTA, BBN, and CMB observations*}},
  \href{https://doi.org/10.1088/1674-1137/ad79d5}{\emph{Chin. Phys. C}
  {\bfseries 48} (2024) 125105},
  [\href{https://arxiv.org/abs/2307.13574}{{\ttfamily 2307.13574}}].

\bibitem{Carr:2021bzv}
B.~Carr and F.~Kuhnel, \emph{{Primordial black holes as dark matter
  candidates}},
  \href{https://doi.org/10.21468/SciPostPhysLectNotes.48}{\emph{SciPost Phys.
  Lect. Notes} {\bfseries 48} (2022) 1},
  [\href{https://arxiv.org/abs/2110.02821}{{\ttfamily 2110.02821}}].

\bibitem{Zhang:2023zmb}
C.~Zhang and X.~Zhang, \emph{{Magnetic monopole meets primordial black hole: an
  extended analysis}},
  \href{https://doi.org/10.1140/epjc/s10052-024-12383-8}{\emph{Eur. Phys. J. C}
  {\bfseries 84} (2024) 100},
  [\href{https://arxiv.org/abs/2308.07166}{{\ttfamily 2308.07166}}].

\bibitem{Yang:2024vij}
C.~Yang, S.~Wang, M.-L. Zhao and X.~Zhang, \emph{{Search for the Hawking
  radiation of primordial black holes: prospective sensitivity of LHAASO}},
  \href{https://doi.org/10.1088/1475-7516/2024/10/083}{\emph{JCAP} {\bfseries
  10} (2024) 083}, [\href{https://arxiv.org/abs/2408.10897}{{\ttfamily
  2408.10897}}].

\bibitem{Yang:2025uvf}
C.~Yang and X.~Zhang, \emph{{Prospective sensitivity of CTA on detection of
  evaporating primordial black holes}},
  \href{https://arxiv.org/abs/2504.17478}{{\ttfamily 2504.17478}}.

\bibitem{Yang:2024pfb}
C.~Yang and X.~Zhang, \emph{{Constraining primordial black hole abundance with
  Insight-HXMT}},  \href{https://arxiv.org/abs/2412.09297}{{\ttfamily
  2412.09297}}.

\bibitem{Zhao:2024yus}
M.-L. Zhao, S.~Wang and X.~Zhang, \emph{{Prospects for detecting the dark
  matter particles and primordial black holes with the Hongmeng mission using
  the 21 cm global spectrum at cosmic dawn}},
  \href{https://arxiv.org/abs/2412.19257}{{\ttfamily 2412.19257}}.

\bibitem{Escriva:2022duf}
A.~Escriv\`a, F.~Kuhnel and Y.~Tada, \emph{{Primordial Black Holes}},
  \href{https://arxiv.org/abs/2211.05767}{{\ttfamily 2211.05767}}.

\bibitem{Carr:2020xqk}
B.~Carr and F.~Kuhnel, \emph{{Primordial Black Holes as Dark Matter: Recent
  Developments}},
  \href{https://doi.org/10.1146/annurev-nucl-050520-125911}{\emph{Ann. Rev.
  Nucl. Part. Sci.} {\bfseries 70} (2020) 355--394},
  [\href{https://arxiv.org/abs/2006.02838}{{\ttfamily 2006.02838}}].

\bibitem{Liu:2021svg}
J.~Liu, L.~Bian, R.-G. Cai, Z.-K. Guo and S.-J. Wang, \emph{{Primordial black
  hole production during first-order phase transitions}},
  \href{https://doi.org/10.1103/PhysRevD.105.L021303}{\emph{Phys. Rev. D}
  {\bfseries 105} (2022) L021303},
  [\href{https://arxiv.org/abs/2106.05637}{{\ttfamily 2106.05637}}].

\bibitem{He:2022amv}
S.~He, L.~Li, Z.~Li and S.-J. Wang, \emph{{Gravitational waves and primordial
  black hole productions from gluodynamics by holography}},
  \href{https://doi.org/10.1007/s11433-023-2293-2}{\emph{Sci. China Phys. Mech.
  Astron.} {\bfseries 67} (2024) 240411},
  [\href{https://arxiv.org/abs/2210.14094}{{\ttfamily 2210.14094}}].

\bibitem{Pi:2021dft}
S.~Pi and M.~Sasaki, \emph{{Primordial black hole formation in nonminimal
  curvaton scenarios}},
  \href{https://doi.org/10.1103/PhysRevD.108.L101301}{\emph{Phys. Rev. D}
  {\bfseries 108} (2023) L101301},
  [\href{https://arxiv.org/abs/2112.12680}{{\ttfamily 2112.12680}}].

\bibitem{Fu:2019ttf}
C.~Fu, P.~Wu and H.~Yu, \emph{{Primordial Black Holes from Inflation with
  Nonminimal Derivative Coupling}},
  \href{https://doi.org/10.1103/PhysRevD.100.063532}{\emph{Phys. Rev. D}
  {\bfseries 100} (2019) 063532},
  [\href{https://arxiv.org/abs/1907.05042}{{\ttfamily 1907.05042}}].

\bibitem{Di:2017ndc}
H.~Di and Y.~Gong, \emph{{Primordial black holes and second order gravitational
  waves from ultra-slow-roll inflation}},
  \href{https://doi.org/10.1088/1475-7516/2018/07/007}{\emph{JCAP} {\bfseries
  07} (2018) 007}, [\href{https://arxiv.org/abs/1707.09578}{{\ttfamily
  1707.09578}}].

\bibitem{Yi:2020kmq}
Z.~Yi, Y.~Gong, B.~Wang and Z.-h. Zhu, \emph{{Primordial black holes and
  secondary gravitational waves from the Higgs field}},
  \href{https://doi.org/10.1103/PhysRevD.103.063535}{\emph{Phys. Rev. D}
  {\bfseries 103} (2021) 063535},
  [\href{https://arxiv.org/abs/2007.09957}{{\ttfamily 2007.09957}}].

\bibitem{Gu:2022pbo}
B.-M. Gu, F.-W. Shu, K.~Yang and Y.-P. Zhang, \emph{{Primordial black holes
  from an inflationary potential valley}},
  \href{https://doi.org/10.1103/PhysRevD.107.023519}{\emph{Phys. Rev. D}
  {\bfseries 107} (2023) 023519},
  [\href{https://arxiv.org/abs/2207.09968}{{\ttfamily 2207.09968}}].

\bibitem{Cai:2018tuh}
Y.-F. Cai, X.~Tong, D.-G. Wang and S.-F. Yan, \emph{{Primordial Black Holes
  from Sound Speed Resonance during Inflation}},
  \href{https://doi.org/10.1103/PhysRevLett.121.081306}{\emph{Phys. Rev. Lett.}
  {\bfseries 121} (2018) 081306},
  [\href{https://arxiv.org/abs/1805.03639}{{\ttfamily 1805.03639}}].

\bibitem{Wang:2016ana}
S.~Wang, Y.-F. Wang, Q.-G. Huang and T.~G.~F. Li, \emph{{Constraints on the
  Primordial Black Hole Abundance from the First Advanced LIGO Observation Run
  Using the Stochastic Gravitational-Wave Background}},
  \href{https://doi.org/10.1103/PhysRevLett.120.191102}{\emph{Phys. Rev. Lett.}
  {\bfseries 120} (2018) 191102},
  [\href{https://arxiv.org/abs/1610.08725}{{\ttfamily 1610.08725}}].

\bibitem{Wang:2022nml}
X.~Wang, Y.-l. Zhang, R.~Kimura and M.~Yamaguchi, \emph{{Reconstruction of
  power spectrum of primordial curvature perturbations on small scales from
  primordial black hole binaries scenario of LIGO/VIRGO detection}},
  \href{https://doi.org/10.1007/s11433-023-2091-x}{\emph{Sci. China Phys. Mech.
  Astron.} {\bfseries 66} (2023) 260462},
  [\href{https://arxiv.org/abs/2209.12911}{{\ttfamily 2209.12911}}].

\bibitem{Huang:2023mwy}
H.-L. Huang and Y.-S. Piao, \emph{{Toward supermassive primordial black holes
  from inflationary bubbles}},
  \href{https://doi.org/10.1103/PhysRevD.110.023501}{\emph{Phys. Rev. D}
  {\bfseries 110} (2024) 023501},
  [\href{https://arxiv.org/abs/2312.11982}{{\ttfamily 2312.11982}}].

\bibitem{Tan:2022lbm}
X.-H. Tan, Y.-J. Yan, T.~Qiu and J.-Q. Xia, \emph{{Searching for the Signal of
  a Primordial Black Hole from CMB Lensing and \ensuremath{\gamma}-Ray
  Emissions}}, \href{https://doi.org/10.3847/2041-8213/ac9668}{\emph{Astrophys.
  J. Lett.} {\bfseries 939} (2022) L15},
  [\href{https://arxiv.org/abs/2209.15222}{{\ttfamily 2209.15222}}].

\bibitem{Friedlander:2022ttk}
A.~Friedlander, K.~J. Mack, S.~Schon, N.~Song and A.~C. Vincent,
  \emph{{Primordial black hole dark matter in the context of extra
  dimensions}}, \href{https://doi.org/10.1103/PhysRevD.105.103508}{\emph{Phys.
  Rev. D} {\bfseries 105} (2022) 103508},
  [\href{https://arxiv.org/abs/2201.11761}{{\ttfamily 2201.11761}}].

\bibitem{Hawking:1975vcx}
S.~W. Hawking, \emph{{Particle Creation by Black Holes}},
  \href{https://doi.org/10.1007/BF02345020}{\emph{Commun. Math. Phys.}
  {\bfseries 43} (1975) 199--220}.

\bibitem{Dvali:2018xpy}
G.~Dvali, \emph{{A Microscopic Model of Holography: Survival by the Burden of
  Memory}},  \href{https://arxiv.org/abs/1810.02336}{{\ttfamily 1810.02336}}.

\bibitem{Dvali:2020wft}
G.~Dvali, L.~Eisemann, M.~Michel and S.~Zell, \emph{{Black hole metamorphosis
  and stabilization by memory burden}},
  \href{https://doi.org/10.1103/PhysRevD.102.103523}{\emph{Phys. Rev. D}
  {\bfseries 102} (2020) 103523},
  [\href{https://arxiv.org/abs/2006.00011}{{\ttfamily 2006.00011}}].

\bibitem{Dvali:2024hsb}
G.~Dvali, J.~S. Valbuena-Berm\'udez and M.~Zantedeschi, \emph{{Memory burden
  effect in black holes and solitons: Implications for PBH}},
  \href{https://doi.org/10.1103/PhysRevD.110.056029}{\emph{Phys. Rev. D}
  {\bfseries 110} (2024) 056029},
  [\href{https://arxiv.org/abs/2405.13117}{{\ttfamily 2405.13117}}].

\bibitem{Cheek:2021odj}
A.~Cheek, L.~Heurtier, Y.~F. Perez-Gonzalez and J.~Turner, \emph{{Primordial
  black hole evaporation and dark matter production. I. Solely Hawking
  radiation}}, \href{https://doi.org/10.1103/PhysRevD.105.015022}{\emph{Phys.
  Rev. D} {\bfseries 105} (2022) 015022},
  [\href{https://arxiv.org/abs/2107.00013}{{\ttfamily 2107.00013}}].

\bibitem{Eroshenko:2021sez}
Y.~N. Eroshenko, \emph{{Spin of primordial black holes in the model with
  collapsing domain walls}},
  \href{https://doi.org/10.1088/1475-7516/2021/12/041}{\emph{JCAP} {\bfseries
  12} (2021) 041}, [\href{https://arxiv.org/abs/2111.03403}{{\ttfamily
  2111.03403}}].

\bibitem{Harada:2017fjm}
T.~Harada, C.-M. Yoo, K.~Kohri and K.-I. Nakao, \emph{{Spins of primordial
  black holes formed in the matter-dominated phase of the Universe}},
  \href{https://doi.org/10.1103/PhysRevD.96.083517}{\emph{Phys. Rev. D}
  {\bfseries 96} (2017) 083517},
  [\href{https://arxiv.org/abs/1707.03595}{{\ttfamily 1707.03595}}].

\bibitem{Kuhnel:2019zbc}
F.~Kuhnel, \emph{{Enhanced Detectability of Spinning Primordial Black Holes}},
  \href{https://doi.org/10.1140/epjc/s10052-020-7807-z}{\emph{Eur. Phys. J. C}
  {\bfseries 80} (2020) 243},
  [\href{https://arxiv.org/abs/1909.04742}{{\ttfamily 1909.04742}}].

\bibitem{Flores:2021tmc}
M.~M. Flores and A.~Kusenko, \emph{{Spins of primordial black holes formed in
  different cosmological scenarios}},
  \href{https://doi.org/10.1103/PhysRevD.104.063008}{\emph{Phys. Rev. D}
  {\bfseries 104} (2021) 063008},
  [\href{https://arxiv.org/abs/2106.03237}{{\ttfamily 2106.03237}}].

\bibitem{Chambers:1997ai}
C.~M. Chambers, W.~A. Hiscock and B.~Taylor, \emph{{Spinning down a black hole
  with scalar fields}},
  \href{https://doi.org/10.1103/PhysRevLett.78.3249}{\emph{Phys. Rev. Lett.}
  {\bfseries 78} (1997) 3249--3251},
  [\href{https://arxiv.org/abs/gr-qc/9703018}{{\ttfamily gr-qc/9703018}}].

\bibitem{Calza:2021czr}
M.~Calz\`a, J.~March-Russell and J.~G. Rosa, \emph{{Evaporating Primordial
  Black Holes, the String Axiverse, and Hot Dark Radiation}},
  \href{https://doi.org/10.1103/PhysRevLett.133.261003}{\emph{Phys. Rev. Lett.}
  {\bfseries 133} (2024) 261003},
  [\href{https://arxiv.org/abs/2110.13602}{{\ttfamily 2110.13602}}].

\bibitem{Calza:2023rjt}
M.~Calz\`a, J.~a.~G. Rosa and F.~Serrano, \emph{{Primordial black hole
  superradiance and evaporation in the string axiverse}},
  \href{https://doi.org/10.1007/JHEP05(2024)140}{\emph{JHEP} {\bfseries 05}
  (2024) 140}, [\href{https://arxiv.org/abs/2306.09430}{{\ttfamily
  2306.09430}}].

\bibitem{Arvanitaki:2009fg}
A.~Arvanitaki, S.~Dimopoulos, S.~Dubovsky, N.~Kaloper and J.~March-Russell,
  \emph{{String Axiverse}},
  \href{https://doi.org/10.1103/PhysRevD.81.123530}{\emph{Phys. Rev. D}
  {\bfseries 81} (2010) 123530},
  [\href{https://arxiv.org/abs/0905.4720}{{\ttfamily 0905.4720}}].

\bibitem{Berti:2009kk}
E.~Berti, V.~Cardoso and A.~O. Starinets, \emph{{Quasinormal modes of black
  holes and black branes}},
  \href{https://doi.org/10.1088/0264-9381/26/16/163001}{\emph{Class. Quant.
  Grav.} {\bfseries 26} (2009) 163001},
  [\href{https://arxiv.org/abs/0905.2975}{{\ttfamily 0905.2975}}].

\bibitem{Arvanitaki:2010sy}
A.~Arvanitaki and S.~Dubovsky, \emph{{Exploring the String Axiverse with
  Precision Black Hole Physics}},
  \href{https://doi.org/10.1103/PhysRevD.83.044026}{\emph{Phys. Rev. D}
  {\bfseries 83} (2011) 044026},
  [\href{https://arxiv.org/abs/1004.3558}{{\ttfamily 1004.3558}}].

\bibitem{Brito:2013wya}
R.~Brito, V.~Cardoso and P.~Pani, \emph{{Massive spin-2 fields on black hole
  spacetimes: Instability of the Schwarzschild and Kerr solutions and bounds on
  the graviton mass}},
  \href{https://doi.org/10.1103/PhysRevD.88.023514}{\emph{Phys. Rev. D}
  {\bfseries 88} (2013) 023514},
  [\href{https://arxiv.org/abs/1304.6725}{{\ttfamily 1304.6725}}].

\bibitem{Brito:2014wla}
R.~Brito, V.~Cardoso and P.~Pani, \emph{{Black holes as particle detectors:
  evolution of superradiant instabilities}},
  \href{https://doi.org/10.1088/0264-9381/32/13/134001}{\emph{Class. Quant.
  Grav.} {\bfseries 32} (2015) 134001},
  [\href{https://arxiv.org/abs/1411.0686}{{\ttfamily 1411.0686}}].

\bibitem{Arvanitaki:2014wva}
A.~Arvanitaki, M.~Baryakhtar and X.~Huang, \emph{{Discovering the QCD Axion
  with Black Holes and Gravitational Waves}},
  \href{https://doi.org/10.1103/PhysRevD.91.084011}{\emph{Phys. Rev. D}
  {\bfseries 91} (2015) 084011},
  [\href{https://arxiv.org/abs/1411.2263}{{\ttfamily 1411.2263}}].

\bibitem{Li:2014fna}
R.~Li and J.~Zhao, \emph{{Numerical study of superradiant instability for
  charged stringy black hole\textendash{}mirror system}},
  \href{https://doi.org/10.1016/j.physletb.2014.12.007}{\emph{Phys. Lett. B}
  {\bfseries 740} (2015) 317--321},
  [\href{https://arxiv.org/abs/1412.1527}{{\ttfamily 1412.1527}}].

\bibitem{Wang:2015fgp}
M.~Wang and C.~Herdeiro, \emph{{Maxwell perturbations on
  Kerr\textendash{}anti\textendash{}de Sitter black holes: Quasinormal modes,
  superradiant instabilities, and vector clouds}},
  \href{https://doi.org/10.1103/PhysRevD.93.064066}{\emph{Phys. Rev. D}
  {\bfseries 93} (2016) 064066},
  [\href{https://arxiv.org/abs/1512.02262}{{\ttfamily 1512.02262}}].

\bibitem{Huang:2016qnk}
Y.~Huang and D.-J. Liu, \emph{{Scalar clouds and the superradiant instability
  regime of Kerr-Newman black hole}},
  \href{https://doi.org/10.1103/PhysRevD.94.064030}{\emph{Phys. Rev. D}
  {\bfseries 94} (2016) 064030},
  [\href{https://arxiv.org/abs/1606.08913}{{\ttfamily 1606.08913}}].

\bibitem{Endlich:2016jgc}
S.~Endlich and R.~Penco, \emph{{A Modern Approach to Superradiance}},
  \href{https://doi.org/10.1007/JHEP05(2017)052}{\emph{JHEP} {\bfseries 05}
  (2017) 052}, [\href{https://arxiv.org/abs/1609.06723}{{\ttfamily
  1609.06723}}].

\bibitem{Rosa:2017ury}
J.~G. Rosa and T.~W. Kephart, \emph{{Stimulated Axion Decay in Superradiant
  Clouds around Primordial Black Holes}},
  \href{https://doi.org/10.1103/PhysRevLett.120.231102}{\emph{Phys. Rev. Lett.}
  {\bfseries 120} (2018) 231102},
  [\href{https://arxiv.org/abs/1709.06581}{{\ttfamily 1709.06581}}].

\bibitem{Baryakhtar:2017ngi}
M.~Baryakhtar, R.~Lasenby and M.~Teo, \emph{{Black Hole Superradiance
  Signatures of Ultralight Vectors}},
  \href{https://doi.org/10.1103/PhysRevD.96.035019}{\emph{Phys. Rev. D}
  {\bfseries 96} (2017) 035019},
  [\href{https://arxiv.org/abs/1704.05081}{{\ttfamily 1704.05081}}].

\bibitem{Huang:2018qdl}
Y.~Huang, D.-J. Liu, X.-h. Zhai and X.-z. Li, \emph{{Instability for massive
  scalar fields in Kerr-Newman spacetime}},
  \href{https://doi.org/10.1103/PhysRevD.98.025021}{\emph{Phys. Rev. D}
  {\bfseries 98} (2018) 025021},
  [\href{https://arxiv.org/abs/1807.06263}{{\ttfamily 1807.06263}}].

\bibitem{Li:2019tns}
R.~Li, Y.~Zhao, T.~Zi and X.~Chen, \emph{{Superradiance and dynamical evolution
  of a charged scalar field in an asymptotically anti\textendash{}de-Sitter
  dilatonic black hole}},
  \href{https://doi.org/10.1103/PhysRevD.99.084045}{\emph{Phys. Rev. D}
  {\bfseries 99} (2019) 084045}.

\bibitem{Chen:2019fsq}
Y.~Chen, J.~Shu, X.~Xue, Q.~Yuan and Y.~Zhao, \emph{{Probing Axions with Event
  Horizon Telescope Polarimetric Measurements}},
  \href{https://doi.org/10.1103/PhysRevLett.124.061102}{\emph{Phys. Rev. Lett.}
  {\bfseries 124} (2020) 061102},
  [\href{https://arxiv.org/abs/1905.02213}{{\ttfamily 1905.02213}}].

\bibitem{Zhang:2020sjh}
C.-Y. Zhang, S.-J. Zhang, P.-C. Li and M.~Guo, \emph{{Superradiance and
  stability of the regularized 4D charged Einstein-Gauss-Bonnet black hole}},
  \href{https://doi.org/10.1007/JHEP08(2020)105}{\emph{JHEP} {\bfseries 08}
  (2020) 105}, [\href{https://arxiv.org/abs/2004.03141}{{\ttfamily
  2004.03141}}].

\bibitem{Liu:2020evp}
P.~Liu, C.~Niu and C.-Y. Zhang, \emph{{Instability of regularized 4D charged
  Einstein-Gauss-Bonnet de-Sitter black holes}},
  \href{https://doi.org/10.1088/1674-1137/abcd2d}{\emph{Chin. Phys. C}
  {\bfseries 45} (2021) 025104},
  [\href{https://arxiv.org/abs/2004.10620}{{\ttfamily 2004.10620}}].

\bibitem{Brito:2020lup}
R.~Brito, S.~Grillo and P.~Pani, \emph{{Black Hole Superradiant Instability
  from Ultralight Spin-2 Fields}},
  \href{https://doi.org/10.1103/PhysRevLett.124.211101}{\emph{Phys. Rev. Lett.}
  {\bfseries 124} (2020) 211101},
  [\href{https://arxiv.org/abs/2002.04055}{{\ttfamily 2002.04055}}].

\bibitem{Baryakhtar:2020gao}
M.~Baryakhtar, M.~Galanis, R.~Lasenby and O.~Simon, \emph{{Black hole
  superradiance of self-interacting scalar fields}},
  \href{https://doi.org/10.1103/PhysRevD.103.095019}{\emph{Phys. Rev. D}
  {\bfseries 103} (2021) 095019},
  [\href{https://arxiv.org/abs/2011.11646}{{\ttfamily 2011.11646}}].

\bibitem{Mai:2021yny}
Z.-F. Mai, R.-Q. Yang and H.~Lu, \emph{{Superradiant instability of extremal
  black holes in STU supergravity}},
  \href{https://doi.org/10.1103/PhysRevD.105.024070}{\emph{Phys. Rev. D}
  {\bfseries 105} (2022) 024070},
  [\href{https://arxiv.org/abs/2110.14942}{{\ttfamily 2110.14942}}].

\bibitem{Xie:2022uvp}
N.~Xie and F.~P. Huang, \emph{{Imprints of ultralight axions on the
  gravitational wave and pulsar timing measurement}},
  \href{https://doi.org/10.1007/s11433-023-2172-7}{\emph{Sci. China Phys. Mech.
  Astron.} {\bfseries 67} (2024) 210411},
  [\href{https://arxiv.org/abs/2207.11145}{{\ttfamily 2207.11145}}].

\bibitem{Bao:2022hew}
S.~Bao, Q.~Xu and H.~Zhang, \emph{{Improved analytic solution of black hole
  superradiance}},
  \href{https://doi.org/10.1103/PhysRevD.106.064016}{\emph{Phys. Rev. D}
  {\bfseries 106} (2022) 064016},
  [\href{https://arxiv.org/abs/2201.10941}{{\ttfamily 2201.10941}}].

\bibitem{Siemonsen:2022ivj}
N.~Siemonsen, C.~Mondino, D.~Egana-Ugrinovic, J.~Huang, M.~Baryakhtar and W.~E.
  East, \emph{{Dark photon superradiance: Electrodynamics and multimessenger
  signals}}, \href{https://doi.org/10.1103/PhysRevD.107.075025}{\emph{Phys.
  Rev. D} {\bfseries 107} (2023) 075025},
  [\href{https://arxiv.org/abs/2212.09772}{{\ttfamily 2212.09772}}].

\bibitem{Cheng:2022jsw}
L.-d. Cheng, H.~Zhang and S.-s. Bao, \emph{{Constraints on an axionlike
  particle from black hole spin superradiance}},
  \href{https://doi.org/10.1103/PhysRevD.107.063021}{\emph{Phys. Rev. D}
  {\bfseries 107} (2023) 063021},
  [\href{https://arxiv.org/abs/2201.11338}{{\ttfamily 2201.11338}}].

\bibitem{Zhou:2023sps}
L.~Zhou, R.~Brito, Z.-F. Mai and L.~Shao, \emph{{Superradiant instabilities of
  massive bosons around exotic compact objects}},
  \href{https://doi.org/10.1103/PhysRevD.108.103025}{\emph{Phys. Rev. D}
  {\bfseries 108} (2023) 103025},
  [\href{https://arxiv.org/abs/2308.03091}{{\ttfamily 2308.03091}}].

\bibitem{Jia:2023see}
N.~Jia, Y.-D. Guo, G.-R. Liang, Z.-F. Mai and X.~Zhang, \emph{{Superradiant
  growth anomaly magnification in evolution of vector bosonic condensates
  bounded by a Kerr black hole with near-horizon reflection}},
  \href{https://doi.org/10.1007/s11433-024-2602-0}{\emph{Sci. China Phys. Mech.
  Astron.} {\bfseries 68} (2025) 240411},
  [\href{https://arxiv.org/abs/2309.05108}{{\ttfamily 2309.05108}}].

\bibitem{Bao:2023xna}
S.-S. Bao, Q.-X. Xu and H.~Zhang, \emph{{Next-to-leading-order solution to
  Kerr-Newman black hole superradiance}},
  \href{https://doi.org/10.1103/PhysRevD.107.064037}{\emph{Phys. Rev. D}
  {\bfseries 107} (2023) 064037},
  [\href{https://arxiv.org/abs/2301.05317}{{\ttfamily 2301.05317}}].

\bibitem{Yang:2023vwm}
J.~Yang and F.~P. Huang, \emph{{Gravitational waves from axions annihilation
  through quantum field theory}},
  \href{https://doi.org/10.1103/PhysRevD.108.103002}{\emph{Phys. Rev. D}
  {\bfseries 108} (2023) 103002},
  [\href{https://arxiv.org/abs/2306.12375}{{\ttfamily 2306.12375}}].

\bibitem{Yang:2023aak}
J.~Yang, N.~Xie and F.~P. Huang, \emph{{Implication of nano-Hertz stochastic
  gravitational wave background on ultralight axion particles}},
  \href{https://doi.org/10.1088/1475-7516/2024/11/045}{\emph{JCAP} {\bfseries
  11} (2024) 045}, [\href{https://arxiv.org/abs/2306.17113}{{\ttfamily
  2306.17113}}].

\bibitem{Dai:2023ewf}
D.-C. Dai and D.~Stojkovic, \emph{{Separating the superradiant emission from
  the Hawking radiation from a rotating black hole}},
  \href{https://doi.org/10.1016/j.physletb.2023.138056}{\emph{Phys. Lett. B}
  {\bfseries 843} (2023) 138056},
  [\href{https://arxiv.org/abs/2306.17423}{{\ttfamily 2306.17423}}].

\bibitem{Dai:2023zcj}
D.-C. Dai and D.~Stojkovic, \emph{{Shedding new light on the absence of
  fermionic superradiance and the maximal infalling rate of fermions into a
  black hole}}, \href{https://doi.org/10.1103/PhysRevD.108.084024}{\emph{Phys.
  Rev. D} {\bfseries 108} (2023) 084024},
  [\href{https://arxiv.org/abs/2309.13511}{{\ttfamily 2309.13511}}].

\bibitem{Chu:2024iie}
X.-h. Chu, Y.-q. Chu, S.-s. Bao and H.~Zhang, \emph{{Next-to-leading order
  corrections to scalar perturbations of Kerr-anti\textendash{}de Sitter black
  holes}}, \href{https://doi.org/10.1103/PhysRevD.111.043039}{\emph{Phys. Rev.
  D} {\bfseries 111} (2025) 043039},
  [\href{https://arxiv.org/abs/2411.09980}{{\ttfamily 2411.09980}}].

\bibitem{Yu:2024mye}
C.~Yu, X.~Zhang, S.~Kazempour and S.~Sun, \emph{{Superradiance in acoustic
  black hole}},  \href{https://arxiv.org/abs/2412.20890}{{\ttfamily
  2412.20890}}.

\bibitem{Xie:2025npy}
N.~Xie and F.~P. Huang, \emph{{The self-interaction effects on the Kerr black
  hole superradiance and their observational implications}},
  \href{https://arxiv.org/abs/2503.10347}{{\ttfamily 2503.10347}}.

\bibitem{Brito:2015oca}
R.~Brito, V.~Cardoso and P.~Pani, \emph{{Superradiance}: {New Frontiers in
  Black Hole Physics}},
  \href{https://doi.org/10.1007/978-3-319-19000-6}{\emph{Lect. Notes Phys.}
  {\bfseries 906} (2015) pp.1--237},
  [\href{https://arxiv.org/abs/1501.06570}{{\ttfamily 1501.06570}}].

\bibitem{Bernal:2022oha}
N.~Bernal, Y.~F. Perez-Gonzalez and Y.~Xu, \emph{{Superradiant production of
  heavy dark matter from primordial black holes}},
  \href{https://doi.org/10.1103/PhysRevD.106.015020}{\emph{Phys. Rev. D}
  {\bfseries 106} (2022) 015020},
  [\href{https://arxiv.org/abs/2205.11522}{{\ttfamily 2205.11522}}].

\bibitem{Cheek:2021cfe}
A.~Cheek, L.~Heurtier, Y.~F. Perez-Gonzalez and J.~Turner, \emph{{Primordial
  black hole evaporation and dark matter production. II. Interplay with the
  freeze-in or freeze-out mechanism}},
  \href{https://doi.org/10.1103/PhysRevD.105.015023}{\emph{Phys. Rev. D}
  {\bfseries 105} (2022) 015023},
  [\href{https://arxiv.org/abs/2107.00016}{{\ttfamily 2107.00016}}].

\bibitem{Bernal:2020ili}
N.~Bernal and O.~Zapata, \emph{{Gravitational dark matter production:
  primordial black holes and UV freeze-in}},
  \href{https://doi.org/10.1016/j.physletb.2021.136129}{\emph{Phys. Lett. B}
  {\bfseries 815} (2021) 136129},
  [\href{https://arxiv.org/abs/2011.02510}{{\ttfamily 2011.02510}}].

\bibitem{Bernal:2018qlk}
N.~Bernal, M.~Dutra, Y.~Mambrini, K.~Olive, M.~Peloso and M.~Pierre,
  \emph{{Spin-2 Portal Dark Matter}},
  \href{https://doi.org/10.1103/PhysRevD.97.115020}{\emph{Phys. Rev. D}
  {\bfseries 97} (2018) 115020},
  [\href{https://arxiv.org/abs/1803.01866}{{\ttfamily 1803.01866}}].

\bibitem{Tang:2017hvq}
Y.~Tang and Y.-L. Wu, \emph{{On Thermal Gravitational Contribution to Particle
  Production and Dark Matter}},
  \href{https://doi.org/10.1016/j.physletb.2017.10.034}{\emph{Phys. Lett. B}
  {\bfseries 774} (2017) 676--681},
  [\href{https://arxiv.org/abs/1708.05138}{{\ttfamily 1708.05138}}].

\bibitem{Garny:2017kha}
M.~Garny, A.~Palessandro, M.~Sandora and M.~S. Sloth, \emph{{Theory and
  Phenomenology of Planckian Interacting Massive Particles as Dark Matter}},
  \href{https://doi.org/10.1088/1475-7516/2018/02/027}{\emph{JCAP} {\bfseries
  02} (2018) 027}, [\href{https://arxiv.org/abs/1709.09688}{{\ttfamily
  1709.09688}}].

\bibitem{Garny:2015sjg}
M.~Garny, M.~Sandora and M.~S. Sloth, \emph{{Planckian Interacting Massive
  Particles as Dark Matter}},
  \href{https://doi.org/10.1103/PhysRevLett.116.101302}{\emph{Phys. Rev. Lett.}
  {\bfseries 116} (2016) 101302},
  [\href{https://arxiv.org/abs/1511.03278}{{\ttfamily 1511.03278}}].

\bibitem{Carr:1974nx}
B.~J. Carr and S.~W. Hawking, \emph{{Black holes in the early Universe}},
  \href{https://doi.org/10.1093/mnras/168.2.399}{\emph{Mon. Not. Roy. Astron.
  Soc.} {\bfseries 168} (1974) 399--415}.

\bibitem{Carr:1975qj}
B.~J. Carr, \emph{{The Primordial black hole mass spectrum}},
  \href{https://doi.org/10.1086/153853}{\emph{Astrophys. J.} {\bfseries 201}
  (1975) 1--19}.

\bibitem{Hawking:1982ga}
S.~W. Hawking, I.~G. Moss and J.~M. Stewart, \emph{{Bubble Collisions in the
  Very Early Universe}},
  \href{https://doi.org/10.1103/PhysRevD.26.2681}{\emph{Phys. Rev. D}
  {\bfseries 26} (1982) 2681}.

\bibitem{Moss:1994iq}
I.~G. Moss, \emph{{Singularity formation from colliding bubbles}},
  \href{https://doi.org/10.1103/PhysRevD.50.676}{\emph{Phys. Rev. D} {\bfseries
  50} (1994) 676--681}.

\bibitem{Khlopov:1999ys}
M.~Y. Khlopov, R.~V. Konoplich, S.~G. Rubin and A.~S. Sakharov, \emph{{First
  order phase transitions as a source of black holes in the early universe}},
  {\emph{Grav. Cosmol.} {\bfseries 2} (1999) S1},
  [\href{https://arxiv.org/abs/hep-ph/9912422}{{\ttfamily hep-ph/9912422}}].

\bibitem{Sato:1981bf}
K.~Sato, M.~Sasaki, H.~Kodama and K.-i. Maeda, \emph{{Creation of Wormholes by
  First Order Phase Transition of a Vacuum in the Early Universe}},
  \href{https://doi.org/10.1143/PTP.65.1443}{\emph{Prog. Theor. Phys.}
  {\bfseries 65} (1981) 1443}.

\bibitem{Maeda:1981gw}
K.-i. Maeda, K.~Sato, M.~Sasaki and H.~Kodama, \emph{{Creation of De
  Sitter-schwarzschild Wormholes by a Cosmological First Order Phase
  Transition}}, \href{https://doi.org/10.1016/0370-2693(82)91151-0}{\emph{Phys.
  Lett. B} {\bfseries 108} (1982) 98--102}.

\bibitem{Sato:1981gv}
K.~Sato, H.~Kodama, M.~Sasaki and K.-i. Maeda, \emph{{Multiproduction of
  Universes by First Order Phase Transition of a Vacuum}},
  \href{https://doi.org/10.1016/0370-2693(82)91152-2}{\emph{Phys. Lett. B}
  {\bfseries 108} (1982) 103--107}.

\bibitem{Kodama:1981gu}
H.~Kodama, M.~Sasaki, K.~Sato and K.-i. Maeda, \emph{{Fate of Wormholes Created
  by First Order Phase Transition in the Early Universe}},
  \href{https://doi.org/10.1143/PTP.66.2052}{\emph{Prog. Theor. Phys.}
  {\bfseries 66} (1981) 2052}.

\bibitem{Kodama:1982sf}
H.~Kodama, M.~Sasaki and K.~Sato, \emph{{Abundance of Primordial Holes Produced
  by Cosmological First Order Phase Transition}},
  \href{https://doi.org/10.1143/PTP.68.1979}{\emph{Prog. Theor. Phys.}
  {\bfseries 68} (1982) 1979}.

\bibitem{Hsu:1990fg}
S.~D.~H. Hsu, \emph{{Black Holes From Extended Inflation}},
  \href{https://doi.org/10.1016/0370-2693(90)90717-K}{\emph{Phys. Lett. B}
  {\bfseries 251} (1990) 343--348}.

\bibitem{Hashino:2021qoq}
K.~Hashino, S.~Kanemura and T.~Takahashi, \emph{{Primordial black holes as a
  probe of strongly first-order electroweak phase transition}},
  \href{https://doi.org/10.1016/j.physletb.2022.137261}{\emph{Phys. Lett. B}
  {\bfseries 833} (2022) 137261},
  [\href{https://arxiv.org/abs/2111.13099}{{\ttfamily 2111.13099}}].

\bibitem{Kawana:2022olo}
K.~Kawana, T.~Kim and P.~Lu, \emph{{PBH formation from overdensities in delayed
  vacuum transitions}},
  \href{https://doi.org/10.1103/PhysRevD.108.103531}{\emph{Phys. Rev. D}
  {\bfseries 108} (2023) 103531},
  [\href{https://arxiv.org/abs/2212.14037}{{\ttfamily 2212.14037}}].

\bibitem{Lewicki:2023ioy}
M.~Lewicki, P.~Toczek and V.~Vaskonen, \emph{{Primordial black holes from
  strong first-order phase transitions}},
  \href{https://doi.org/10.1007/JHEP09(2023)092}{\emph{JHEP} {\bfseries 09}
  (2023) 092}, [\href{https://arxiv.org/abs/2305.04924}{{\ttfamily
  2305.04924}}].

\bibitem{Gehrman:2023esa}
T.~C. Gehrman, B.~Shams Es~Haghi, K.~Sinha and T.~Xu, \emph{{The primordial
  black holes that disappeared: connections to dark matter and MHz-GHz
  gravitational Waves}},
  \href{https://doi.org/10.1088/1475-7516/2023/10/001}{\emph{JCAP} {\bfseries
  10} (2023) 001}, [\href{https://arxiv.org/abs/2304.09194}{{\ttfamily
  2304.09194}}].

\bibitem{Gouttenoire:2023naa}
Y.~Gouttenoire and T.~Volansky, \emph{{Primordial black holes from supercooled
  phase transitions}},
  \href{https://doi.org/10.1103/PhysRevD.110.043514}{\emph{Phys. Rev. D}
  {\bfseries 110} (2024) 043514},
  [\href{https://arxiv.org/abs/2305.04942}{{\ttfamily 2305.04942}}].

\bibitem{Rubin:2000dq}
S.~G. Rubin, M.~Y. Khlopov and A.~S. Sakharov, \emph{{Primordial black holes
  from nonequilibrium second order phase transition}}, {\emph{Grav. Cosmol.}
  {\bfseries 6} (2000) 51--58},
  [\href{https://arxiv.org/abs/hep-ph/0005271}{{\ttfamily hep-ph/0005271}}].

\bibitem{Rubin:2001yw}
S.~G. Rubin, A.~S. Sakharov and M.~Y. Khlopov, \emph{{The Formation of primary
  galactic nuclei during phase transitions in the early universe}},
  \href{https://doi.org/10.1134/1.1385631}{\emph{J. Exp. Theor. Phys.}
  {\bfseries 91} (2001) 921--929},
  [\href{https://arxiv.org/abs/hep-ph/0106187}{{\ttfamily hep-ph/0106187}}].

\bibitem{Dokuchaev:2004kr}
V.~Dokuchaev, Y.~Eroshenko and S.~Rubin, \emph{{Quasars formation around
  clusters of primordial black holes}}, {\emph{Grav. Cosmol.} {\bfseries 11}
  (2005) 99--104}, [\href{https://arxiv.org/abs/astro-ph/0412418}{{\ttfamily
  astro-ph/0412418}}].

\bibitem{Linde:1990yj}
A.~D. Linde and D.~H. Lyth, \emph{{Axionic domain wall production during
  inflation}}, \href{https://doi.org/10.1016/0370-2693(90)90613-B}{\emph{Phys.
  Lett. B} {\bfseries 246} (1990) 353--358}.

\bibitem{Carr:2020gox}
B.~Carr, K.~Kohri, Y.~Sendouda and J.~Yokoyama, \emph{{Constraints on
  primordial black holes}},
  \href{https://doi.org/10.1088/1361-6633/ac1e31}{\emph{Rept. Prog. Phys.}
  {\bfseries 84} (2021) 116902},
  [\href{https://arxiv.org/abs/2002.12778}{{\ttfamily 2002.12778}}].

\bibitem{Domenech:2020ssp}
G.~Dom{\`e}nech, C.~Lin and M.~Sasaki, \emph{{Gravitational wave constraints on
  the primordial black hole dominated early universe}},
  \href{https://doi.org/10.1088/1475-7516/2021/11/E01}{\emph{JCAP} {\bfseries
  04} (2021) 062}, [\href{https://arxiv.org/abs/2012.08151}{{\ttfamily
  2012.08151}}].

\bibitem{Planck:2018jri}
{\scshape Planck} collaboration, Y.~Akrami et~al., \emph{{Planck 2018 results.
  X. Constraints on inflation}},
  \href{https://doi.org/10.1051/0004-6361/201833887}{\emph{Astron. Astrophys.}
  {\bfseries 641} (2020) A10},
  [\href{https://arxiv.org/abs/1807.06211}{{\ttfamily 1807.06211}}].

\bibitem{Papanikolaou:2020qtd}
T.~Papanikolaou, V.~Vennin and D.~Langlois, \emph{{Gravitational waves from a
  universe filled with primordial black holes}},
  \href{https://doi.org/10.1088/1475-7516/2021/03/053}{\emph{JCAP} {\bfseries
  03} (2021) 053}, [\href{https://arxiv.org/abs/2010.11573}{{\ttfamily
  2010.11573}}].

\bibitem{Papanikolaou:2022chm}
T.~Papanikolaou, \emph{{Gravitational waves induced from primordial black hole
  fluctuations: the~effect of an extended mass function}},
  \href{https://doi.org/10.1088/1475-7516/2022/10/089}{\emph{JCAP} {\bfseries
  10} (2022) 089}, [\href{https://arxiv.org/abs/2207.11041}{{\ttfamily
  2207.11041}}].

\bibitem{Domenech:2024wao}
G.~Dom\`enech and J.~Tr\"ankle, \emph{{From formation to evaporation: Induced
  gravitational wave probes of the primordial black hole reheating scenario}},
  \href{https://doi.org/10.1103/PhysRevD.111.063528}{\emph{Phys. Rev. D}
  {\bfseries 111} (2025) 063528},
  [\href{https://arxiv.org/abs/2409.12125}{{\ttfamily 2409.12125}}].

\bibitem{Caprini:2018mtu}
C.~Caprini and D.~G. Figueroa, \emph{{Cosmological Backgrounds of Gravitational
  Waves}}, \href{https://doi.org/10.1088/1361-6382/aac608}{\emph{Class. Quant.
  Grav.} {\bfseries 35} (2018) 163001},
  [\href{https://arxiv.org/abs/1801.04268}{{\ttfamily 1801.04268}}].

\bibitem{Dong:2015yjs}
R.~Dong, W.~H. Kinney and D.~Stojkovic, \emph{{Gravitational wave production by
  Hawking radiation from rotating primordial black holes}},
  \href{https://doi.org/10.1088/1475-7516/2016/10/034}{\emph{JCAP} {\bfseries
  10} (2016) 034}, [\href{https://arxiv.org/abs/1511.05642}{{\ttfamily
  1511.05642}}].

\bibitem{Page:1976df}
D.~N. Page, \emph{{Particle Emission Rates from a Black Hole: Massless
  Particles from an Uncharged, Nonrotating Hole}},
  \href{https://doi.org/10.1103/PhysRevD.13.198}{\emph{Phys. Rev. D} {\bfseries
  13} (1976) 198--206}.

\bibitem{Page:1976ki}
D.~N. Page, \emph{{Particle Emission Rates from a Black Hole. 2. Massless
  Particles from a Rotating Hole}},
  \href{https://doi.org/10.1103/PhysRevD.14.3260}{\emph{Phys. Rev. D}
  {\bfseries 14} (1976) 3260--3273}.

\bibitem{Page:1977um}
D.~N. Page, \emph{{Particle Emission Rates from a Black Hole. 3. Charged
  Leptons from a Nonrotating Hole}},
  \href{https://doi.org/10.1103/PhysRevD.16.2402}{\emph{Phys. Rev. D}
  {\bfseries 16} (1977) 2402--2411}.

\bibitem{Chiba:2017rvs}
T.~Chiba and S.~Yokoyama, \emph{{Spin Distribution of Primordial Black Holes}},
  \href{https://doi.org/10.1093/ptep/ptx087}{\emph{PTEP} {\bfseries 2017}
  (2017) 083E01}, [\href{https://arxiv.org/abs/1704.06573}{{\ttfamily
  1704.06573}}].

\bibitem{Mirbabayi:2019uph}
M.~Mirbabayi, A.~Gruzinov and J.~Nore{\~n}a, \emph{{Spin of Primordial Black
  Holes}}, \href{https://doi.org/10.1088/1475-7516/2020/03/017}{\emph{JCAP}
  {\bfseries 03} (2020) 017},
  [\href{https://arxiv.org/abs/1901.05963}{{\ttfamily 1901.05963}}].

\bibitem{He:2019cdb}
M.~He and T.~Suyama, \emph{{Formation threshold of rotating primordial black
  holes}}, \href{https://doi.org/10.1103/PhysRevD.100.063520}{\emph{Phys. Rev.
  D} {\bfseries 100} (2019) 063520},
  [\href{https://arxiv.org/abs/1906.10987}{{\ttfamily 1906.10987}}].

\bibitem{Chongchitnan:2021ehn}
S.~Chongchitnan and J.~Silk, \emph{{Extreme-value statistics of the spin of
  primordial black holes}},
  \href{https://doi.org/10.1103/PhysRevD.104.083018}{\emph{Phys. Rev. D}
  {\bfseries 104} (2021) 083018},
  [\href{https://arxiv.org/abs/2109.12268}{{\ttfamily 2109.12268}}].

\bibitem{Garcia-Bellido:2020pwq}
J.~Garc{\'\i}a-Bellido, J.~F. Nu{\~n}o~Siles and E.~Ruiz~Morales,
  \emph{{Bayesian analysis of the spin distribution of LIGO/Virgo black
  holes}}, \href{https://doi.org/10.1016/j.dark.2021.100791}{\emph{Phys. Dark
  Univ.} {\bfseries 31} (2021) 100791},
  [\href{https://arxiv.org/abs/2010.13811}{{\ttfamily 2010.13811}}].

\bibitem{Harada:2020pzb}
T.~Harada, C.-M. Yoo, K.~Kohri, Y.~Koga and T.~Monobe, \emph{{Spins of
  primordial black holes formed in the radiation-dominated phase of the
  universe: first-order effect}},
  \href{https://doi.org/10.3847/1538-4357/abd9b9}{\emph{Astrophys. J.}
  {\bfseries 908} (2021) 140},
  [\href{https://arxiv.org/abs/2011.00710}{{\ttfamily 2011.00710}}].

\bibitem{Banerjee:2024nkv}
I.~K. Banerjee and T.~Harada, \emph{{Spin of primordial black holes from broad
  power spectrum: radiation dominated universe}},
  \href{https://doi.org/10.1088/1475-7516/2025/05/010}{\emph{JCAP} {\bfseries
  05} (2025) 010}, [\href{https://arxiv.org/abs/2409.06494}{{\ttfamily
  2409.06494}}].

\bibitem{Arbey:2019mbc}
A.~Arbey and J.~Auffinger, \emph{{BlackHawk: A public code for calculating the
  Hawking evaporation spectra of any black hole distribution}},
  \href{https://doi.org/10.1140/epjc/s10052-019-7161-1}{\emph{Eur. Phys. J. C}
  {\bfseries 79} (2019) 693},
  [\href{https://arxiv.org/abs/1905.04268}{{\ttfamily 1905.04268}}].

\bibitem{Baldes:2020nuv}
I.~Baldes, Q.~Decant, D.~C. Hooper and L.~Lopez-Honorez, \emph{{Non-Cold Dark
  Matter from Primordial Black Hole Evaporation}},
  \href{https://doi.org/10.1088/1475-7516/2020/08/045}{\emph{JCAP} {\bfseries
  08} (2020) 045}, [\href{https://arxiv.org/abs/2004.14773}{{\ttfamily
  2004.14773}}].

\bibitem{Baumann:2019eav}
D.~Baumann, H.~S. Chia, J.~Stout and L.~ter Haar, \emph{{The Spectra of
  Gravitational Atoms}},
  \href{https://doi.org/10.1088/1475-7516/2019/12/006}{\emph{JCAP} {\bfseries
  12} (2019) 006}, [\href{https://arxiv.org/abs/1908.10370}{{\ttfamily
  1908.10370}}].

\bibitem{Brito:2017zvb}
R.~Brito, S.~Ghosh, E.~Barausse, E.~Berti, V.~Cardoso, I.~Dvorkin et~al.,
  \emph{{Gravitational wave searches for ultralight bosons with LIGO and
  LISA}}, \href{https://doi.org/10.1103/PhysRevD.96.064050}{\emph{Phys. Rev. D}
  {\bfseries 96} (2017) 064050},
  [\href{https://arxiv.org/abs/1706.06311}{{\ttfamily 1706.06311}}].

\bibitem{Guo:2022mpr}
Y.-d. Guo, S.-s. Bao and H.~Zhang, \emph{{Subdominant modes of the scalar
  superradiant instability and gravitational wave beats}},
  \href{https://doi.org/10.1103/PhysRevD.107.075009}{\emph{Phys. Rev. D}
  {\bfseries 107} (2023) 075009},
  [\href{https://arxiv.org/abs/2212.07186}{{\ttfamily 2212.07186}}].

\bibitem{Guo:2024dqd}
Y.-D. Guo, N.~Jia, S.-S. Bao, H.~Zhang and X.~Zhang, \emph{{Evolution and
  detection of vector superradiant instabilities}},
  \href{https://doi.org/10.1103/PhysRevD.110.083029}{\emph{Phys. Rev. D}
  {\bfseries 110} (2024) 083029},
  [\href{https://arxiv.org/abs/2407.00767}{{\ttfamily 2407.00767}}].

\bibitem{Yoshino:2013ofa}
H.~Yoshino and H.~Kodama, \emph{{Gravitational radiation from an axion cloud
  around a black hole: Superradiant phase}},
  \href{https://doi.org/10.1093/ptep/ptu029}{\emph{PTEP} {\bfseries 2014}
  (2014) 043E02}, [\href{https://arxiv.org/abs/1312.2326}{{\ttfamily
  1312.2326}}].

\bibitem{Siemonsen:2019ebd}
N.~Siemonsen and W.~E. East, \emph{{Gravitational wave signatures of ultralight
  vector bosons from black hole superradiance}},
  \href{https://doi.org/10.1103/PhysRevD.101.024019}{\emph{Phys. Rev. D}
  {\bfseries 101} (2020) 024019},
  [\href{https://arxiv.org/abs/1910.09476}{{\ttfamily 1910.09476}}].

\bibitem{March-Russell:2022zll}
J.~March-Russell and J.~a.~G. Rosa, \emph{{Micro-Bose/Proca dark matter stars
  from black hole superradiance}},
  \href{https://arxiv.org/abs/2205.15277}{{\ttfamily 2205.15277}}.

\bibitem{Giudice:2000ex}
G.~F. Giudice, E.~W. Kolb and A.~Riotto, \emph{{Largest temperature of the
  radiation era and its cosmological implications}},
  \href{https://doi.org/10.1103/PhysRevD.64.023508}{\emph{Phys. Rev. D}
  {\bfseries 64} (2001) 023508},
  [\href{https://arxiv.org/abs/hep-ph/0005123}{{\ttfamily hep-ph/0005123}}].

\bibitem{Aggarwal:2020olq}
N.~Aggarwal et~al., \emph{{Challenges and opportunities of gravitational-wave
  searches at MHz to GHz frequencies}},
  \href{https://doi.org/10.1007/s41114-021-00032-5}{\emph{Living Rev. Rel.}
  {\bfseries 24} (2021) 4}, [\href{https://arxiv.org/abs/2011.12414}{{\ttfamily
  2011.12414}}].

\bibitem{Ghoshal:2023fno}
A.~Ghoshal, Y.~F. Perez-Gonzalez and J.~Turner, \emph{{Superradiant
  leptogenesis}}, \href{https://doi.org/10.1007/JHEP02(2024)113}{\emph{JHEP}
  {\bfseries 02} (2024) 113},
  [\href{https://arxiv.org/abs/2312.06768}{{\ttfamily 2312.06768}}].

\bibitem{Manno:2025dhw}
M.~Manno and D.~Montanino, \emph{{ALPs production from Light Primordial Black
  Holes: the role of Superradiance}},
  \href{https://arxiv.org/abs/2501.14589}{{\ttfamily 2501.14589}}.

\bibitem{Jiang:2025blz}
S.~Jiang and F.~P. Huang, \emph{{Pseudo-Goldstone Dark Matter from Primordial
  Black Holes: Gravitational Wave Signatures and Implications for KM3-230213A
  Event at KM3NeT}},  \href{https://arxiv.org/abs/2503.14332}{{\ttfamily
  2503.14332}}.

\bibitem{Dolan:2007mj}
S.~R. Dolan, \emph{{Instability of the massive Klein-Gordon field on the Kerr
  spacetime}}, \href{https://doi.org/10.1103/PhysRevD.76.084001}{\emph{Phys.
  Rev. D} {\bfseries 76} (2007) 084001},
  [\href{https://arxiv.org/abs/0705.2880}{{\ttfamily 0705.2880}}].

\bibitem{Dolan:2018dqv}
S.~R. Dolan, \emph{{Instability of the Proca field on Kerr spacetime}},
  \href{https://doi.org/10.1103/PhysRevD.98.104006}{\emph{Phys. Rev. D}
  {\bfseries 98} (2018) 104006},
  [\href{https://arxiv.org/abs/1806.01604}{{\ttfamily 1806.01604}}].

\bibitem{Weinberg:1972kfs}
S.~Weinberg, \emph{{Gravitation and Cosmology}: {Principles and Applications of
  the General Theory of Relativity}}.
\newblock John Wiley and Sons, New York, 1972.

\end{thebibliography}\endgroup
\bibliographystyle{JHEP}

\end{document}